\documentclass[%
 reprint,
 superscriptaddress,
 amsmath,
 amssymb,
 amsfonts,
 aps,
 prb,
]{revtex4-2}

\usepackage{graphicx}
\usepackage{dcolumn}
\usepackage{bm}
\usepackage{hyperref}
\usepackage[english]{babel}

\usepackage{xcolor}

\newcommand{\Tc}{T_c}
\newcommand{\FC}{\mathrm{FC}}
\newcommand{\pk}{\mathrm{pk}}
\newcommand{\PL}{\mathrm{PL}}
\newcommand{\mA}{\mathrm{mA}}
\newcommand{\uA}{\mu\mathrm{A}}
\newcommand{\mT}{\mathrm{mT}}
\newcommand{\um}{\mu\mathrm{m}}
\newcommand{\nm}{\mathrm{nm}}
\renewcommand{\vec}{\mathbf}
\renewcommand{\Im}{\mathrm{Im}}
\renewcommand{\Re}{\mathrm{Re}}

\begin{document}

\title{Vortex dynamics induced by scanning SQUID susceptometry}

\author{Logan Bishop-Van Horn}
\altaffiliation{These authors contributed equally.}
\affiliation{Stanford Institute for Materials and Energy Sciences, SLAC National Accelerator Laboratory, Menlo Park, California 94025, USA}
\affiliation{Department of Physics, Stanford University, Stanford, California 94305, USA}

\author{Eli Mueller}
\altaffiliation{These authors contributed equally.}
\affiliation{Stanford Institute for Materials and Energy Sciences, SLAC National Accelerator Laboratory, Menlo Park, California 94025, USA}
\affiliation{Department of Physics, Stanford University, Stanford, California 94305, USA}

\author{Kathryn A. Moler}
\email{kmoler@stanford.edu}
\affiliation{Stanford Institute for Materials and Energy Sciences, SLAC National Accelerator Laboratory, Menlo Park, California 94025, USA}
\affiliation{Department of Physics, Stanford University, Stanford, California 94305, USA}
\affiliation{Department of Applied Physics, Stanford University, Stanford, California 94305, USA}

\date{\today}

\begin{abstract}
We measured the local magnetic response of a niobium thin film by applying a millitesla-scale AC magnetic field using a micron-scale field coil and detecting the response with a micron-scale pickup loop in a scanning superconducting quantum interference device (SQUID) susceptometry measurement. Near the film's critical temperature, we observed a step-like nonlinear and dissipative magnetic response due to the dynamics of a small number of vortex-antivortex pairs induced in the film by the local applied AC field.  We modeled the dynamics of the measurement using a combined two-dimensional London-Maxwell and time-dependent Ginzburg-Landau approach, allowing us to construct a detailed real-space picture of the vortex motion causing the observed dissipative response. This work pushes scanning SQUID susceptometry of two-dimensional superconductors beyond the regime of linear response and lays the foundation for microscopic studies of vortex dynamics and pinning in superconducting devices and more exotic materials systems.
\end{abstract}

\maketitle


\section{Introduction}

One of the most striking phenomena associated with superconductivity is the spontaneous expulsion of both static and time-dependent magnetic fields, the Meissner effect. The Meissner effect is a direct manifestation of the coherence of the superconducting condensate, characterized by a macroscopic complex order parameter $\Psi=|\Psi|e^{i\theta}$. Observing the Meissner effect by measuring the response of superconductors to applied magnetic fields is one of the fundamental methods for probing the superconducting state. Typically one measures a quantity related to the complex AC volume susceptibility, $\chi=\chi'+i\chi''$, by applying an AC magnetic field at angular frequency $\omega$ with a field coil and detecting the sample's magnetic response with a pickup loop~\cite{Fiory1988-zw, Jeanneret1989-ss, Clem1992-ko, Claassen1997-sb, He2016-pz}. The real part $\chi'$ of the AC susceptibility is due to the dissipationless superfluid response of the superconductor and provides a measure of the London penetration depth, $\lambda$, which is in turn related to the superfluid density $n_s=|\Psi|^2\propto\lambda^{-2}$. The imaginary part $\chi''$ is related to energy dissipation~\cite{Clem1991-vq, Goldfard1991-ws, Clem1994-qu}. For temperatures $T$ that are small compared to the critical temperature $\Tc$, $\chi''$ is usually small and $n_s(T)$ is sensitive to the structure of the superconducting gap~\cite{Prozorov2011-vi}. Measurements of $\chi$ as a function of temperature show a peak in the dissipative component $\chi''$ around $\Tc$, the width of which is considered a measure of the homogeneity of the superconductor~\cite{Fiory1988-zw, Bozovic2016-sl}. 

In type-II superconductors, and thin films which can be effectively type-II even if grown from a material that is type-I in the bulk~\cite{Tinkham1964-vo, Fetter1967-ui}, the Meissner state persists up to an effective lower critical field $B_{c1}^\mathrm{eff}$\footnote{We define the effective lower critical field $B_{c1}^\mathrm{eff}$ to be the applied field at which vortices actually begin to penetrate the superconductor. This is an empirical value that depends on the spatial distribution of the applied magnetic field, surface defects, etc. The thermodynamic critical field $B_c<B_{c1}^\mathrm{eff}$ is the field above which a state with one or more vortices has a lower free energy than the vortex-free state, assuming a spatially uniform applied field. Between $B_c$ and $B_{c1}^\mathrm{eff}$, a metastable vortex-free state can exist due to the surface barrier or the free energy barrier to generate a spatially-separated vortex-antivortex pair~\cite{Lemberger2013-ha,Lemberger2013-lu}. $B_{c1}^\mathrm{eff}$ is sometimes called the superheating field~\cite{Matricon1967-oh}.}. Above $B_{c1}^\mathrm{eff}$, magnetic flux can pass through the superconductor in the form of vortices, which are topological defects in the condensate at which the superfluid density $|\Psi|^2$ goes to zero and around which the phase of the order parameter $\theta=\arg(\Psi)$ changes by $2\pi$.
The dissipative component of the AC susceptibility, $\chi''$, arises from the motion of vortices, which experience an oscillating force due to screening currents induced by the applied AC magnetic field~\cite{Clem1991-vq}. AC susceptibility measurements using the two-coil mutual inductance technique~\cite{Fiory1988-zw, Jeanneret1989-ss, Claassen1997-sb} typically employ field coils and pickup loops of diameter $\gtrsim$ 1 mm. As a result, these experiments measure the average response of the film, potentially with contributions from many vortices and from spatial inhomogeneities on submillimeter length scales.

In this work, we measured the low-frequency AC susceptibility of a niobium thin film close to its critical temperature using a superconducting quantum interference device (SQUID) susceptometer with a micron-scale field coil and pickup loop. We observed distinct steps in both the in-phase (superfluid) and out-of-phase (dissipative) components of the susceptibility with increasing temperature and local applied AC field. These steps are clear ``fingerprints'' of the dynamics of a small number of vortices induced in the film by the SQUID susceptometer. We modeled the dynamics of the measurement using a combined two-dimensional (2D) London-Maxwell~\cite{Brandt2004-ew,Brandt2005-wj,Bishop-Van_Horn2022-sy} and time-dependent Ginzburg-Landau (TDGL)~\cite{Watts-Tobin1981-mn, Kramer1978-kb, Jonsson2022-mb, Bishop-Van_Horn2023-wr} approach, allowing us to construct a detailed real-space picture of the vortex motion causing the observed dissipative response.

During one half of the AC cycle, one or more vortex-antivortex pairs are induced in the film if the peak applied field exceeds the effective lower critical field $B_{c1}^\mathrm{eff}$~\footnote{We use the convention that a vortex is associated with a circulating current in the same direction as the instantaneous current in the field coil, and an antivortex has a circulating current in the opposite direction.}. The vortices are pulled toward the center of the field coil, where they are trapped by the local applied field. The antivortices are pushed away from the field coil, causing some of them to exit the film or become pinned far from the field coil so that they no longer participate in the measurement, resulting in a hysteretic magnetic response. The time-dependent spatial distribution of vortices is determined by a competition between the vortex-field coil interaction, the repulsive vortex-vortex interaction, and the attractive vortex-antivortex interaction.

Generating vortices and measuring their dynamics with a micron-scale sensor provides a complementary approach to bulk magnetization, AC susceptibility, and transport measurements. For finite-sized films, these more ``global'' measurements can be dominated by a surface barrier that depends sensitively on the geometry and characteristics of the sample edge~\cite{Bean1964-hv, Hernandez2002-nx, Benfenati2020-mg}. For example, it has been shown that finite-size effects dramatically alter transport characteristics of thin films and Josephson junction arrays near the Berezinskii–Kosterlitz–Thouless (BKT) transition, potentially obscuring the vortex unbinding transition entirely~\cite{Herbert1998-ig, Gurevich2008-vi}.

Thus, in addition to resolving vortex dynamics at the single-vortex level, scanning SQUID susceptometry provides a route to study and potentially minimize the contribution of surface effects in order to more directly probe vortex-antivortex and vortex-defect interactions in 2D superconducting systems. Moreover, the numerical methods presented here allow one to model (in the weak screening limit) vortex dynamics for any spatial distribution of applied field. The methods can be applied to mesoscopic 2D superconducting devices of any geometry, including those with holes, normal metal contacts, edge defects, or spatial inhomogeneity in the critical temperature.

\section{Background}

\subsection{AC losses in type-II superconductors}
\label{sec:ac-loss}

As described in Ref.~\cite{Clem1991-vq}, there are broadly three different mechanisms for loss in type-II superconductors subject to applied AC magnetic fields: (1) viscous flux flow, (2) bulk flux pinning, and (3) surface flux pinning. All three mechanisms are related to the fact that a vortex in the presence of a current experiences a Lorentz force per unit length
\begin{equation}
    \label{eq:lorentz}
    \mathbf{f}=\mathbf{J}_s\times\hat{\mathbf{n}}\Phi_0,
\end{equation}
where $\hat{\mathbf{n}}$ is a unit vector indicating the axis of the vortex core, $\Phi_0=h/2e$ is the superconducting flux quantum, and $\vec{J}_s$ is the supercurrent density. Meissner screening of a time-dependent applied magnetic field, resulting in a time-dependent supercurrent density $\mathbf{J}_s$, will exert a dynamic Lorentz force on a vortex in the superconductor. The vortex will move due to the Lorentz force if it is not strongly pinned, and motion of the normal core of the vortex dissipates energy. For this reason, engineering the pinning landscape to reduce dissipation is one of the key challenges in applied superconductivity~\cite{Eley2021-lu}.

Loss mechanism (1) is analogous to eddy current losses in normal metals, with the normal state resistivity replaced by an effective flux flow resistivity, which depends on the applied magnetic field. Eddy current losses, for which the energy dissipated per AC cycle depends strongly on frequency $\omega$, are negligible at low frequencies and for small samples when $\omega\tau_f\ll1$, where $\tau_f$ is the characteristic time required for changes in the magnetic flux to diffuse through the sample.

Mechanisms (2) and (3) are hysteretic losses, which occur when the flux in the superconductor is out of equilibrium with the applied field due to vortex pinning~\cite{Clem1994-qu}. For these mechanisms, the energy dissipated per AC cycle is independent of frequency. Bulk flux pinning losses, where energy is dissipated as vortices ``hop'' between pinning sites, dominate at low frequency in superconductors with many pinning centers or a large critical current density. Bulk pinning losses also include dissipation arising from the annihilation of vortices with antivortices in the bulk~\cite{Clem1979-tm}.

Surface pinning losses occur when the flux in the superconductor is out of equilibrium with the applied field due to the presence of a surface barrier, an energy barrier that must be overcome for a vortex to enter or leave the superconductor~\cite{Bean1964-hv, Hernandez2002-nx}. The surface barrier arises primarily because the direction of the Meissner screening current flowing near the sample surface (or edge in the case of a thin film in an out-of-plane magnetic field) is opposite the direction of the circulating current around a vortex located inside the superconductor. A detailed Ginzburg-Landau analysis of the surface barrier for uniform applied fields in 2D superconductors, including nonidealities such as surface roughness, is given in Ref.~\cite{Benfenati2020-mg}.

\begin{figure*}
    \centering
    \includegraphics[width=\textwidth]{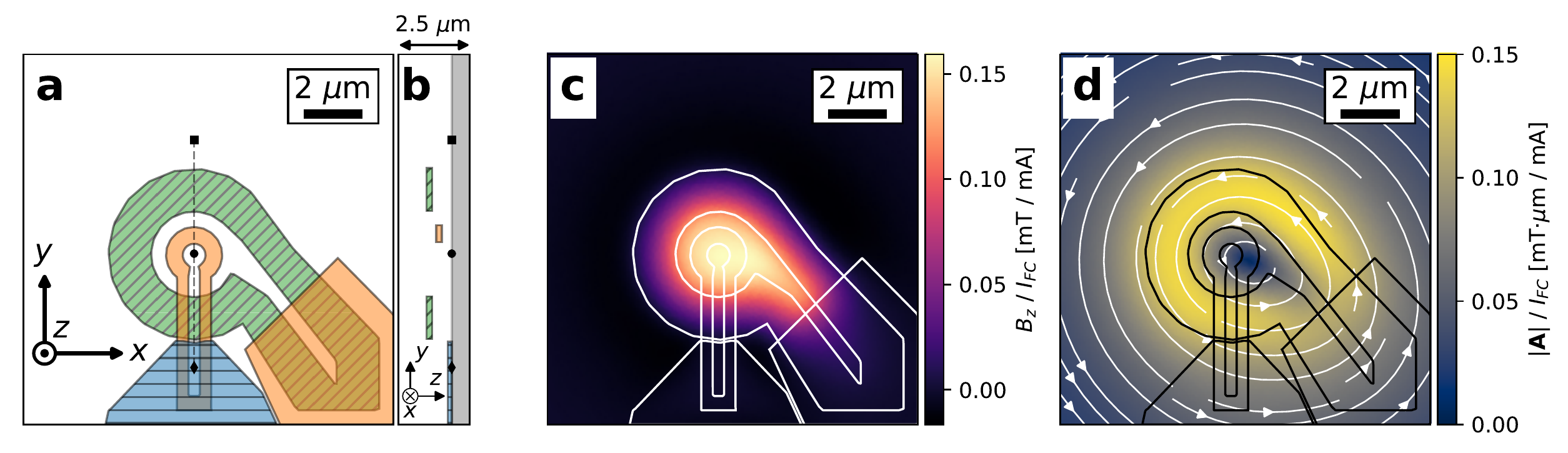}
    \caption{{\bf London-Maxwell simulation of the applied magnetic field and vector potential from the SQUID susceptometer.} ({\bf a}) Modeled geometry of the front field coil-pickup loop pair. A current $I_\FC$ flows counterclockwise in the field coil (green loop, diagonal hatches), and the SQUID measures the flux $\Phi_\PL$ through the front pickup loop (orange loop, no hatches). The blue polygon with horizontal hatches and orange polygon with no hatches are superconducting shields covering the pickup loop and field coil leads, respectively. The field coil, pickup loop, and shields are made of niobium (Nb). ({\bf b}) Layer structure of the SQUID susceptometer, shown along the cut indicated by the dashed gray line in ({\bf a}). The blue wiring layer (135 nm thick) is closest to the sample surface during measurement, the green wiring layer (200 nm thick) is farthest from the surface, and the orange wiring layer (200 nm thick) sits in between. The white region is the SiO${}_2$ insulator of the SQUID chip and the gray region is vacuum between the SQUID and the sample. ({\bf c}) Out-of-plane component of the magnetic field from the field coil per unit field coil current, $B_z/I_\FC$, evaluated at a plane located $z_0=0.5\,\um$ from the surface of the SQUID chip. The maximum value near the center of the field coil is approximately $0.16\,\mT/\mA$. Here, $I_\FC>0$ indicates a counterclockwise circulating current in the field coil. The rest of the SQUID circuit, lying to the south and east as it is drawn here, is well-shielded and does not apply any significant field to the sample. ({\bf d}) In-plane magnetic vector potential per unit field coil current, $\vec{A}/I_\FC=(A_x\hat{\mathbf{x}} + A_y\hat{\mathbf{y}})/I_\FC$, evaluated in the Lorenz gauge at a plane $z_0=0.5\,\um$ below the SQUID. ({\bf c}) and ({\bf d}) are related by $\vec{\nabla}\times\vec{A}=B_z\hat{\mathbf{z}}$. The simulation method is described in Appendix~\ref{sec:london-maxwell} and Ref.~\cite{Bishop-Van_Horn2022-sy}. As shown in ({\bf b}), the layer structure of the SQUID susceptometer is such that, for a given $z_0$, the actual distance from the sample surface to the pickup loop and field coil are approximately $z_0+0.4\,\um$ and $z_0+0.8\,\um$, respectively.}
    \label{fig:squid}
\end{figure*}

For AC susceptibility measurements of thin film superconductors, it is convenient to apply the magnetic field and detect the sample response using coils that are much smaller than the size of the film, both to simplify analysis of the measured magnetic response~\cite{Jeanneret1989-ss, Turneaure1996-ns, Turneaure1998-mo, He2016-pz} and to minimize the extent to which the measurement averages over spatial inhomogeneity. In the limit that the field coil is much smaller than the film and located close to the film surface and far from the film edge, the applied magnetic field at the edge approaches zero and the Meissner screening contribution to the surface energy barrier vanishes. In this case, the onset of vortex-related losses will be dictated by the energy barrier for generating (or ``unbinding'') a vortex-antivortex pair within the film~\cite{Lemberger2016-pb}.

\subsection{Scanning SQUID susceptometry}
\label{sec:susceptometry}

In scanning SQUID susceptometry, a micron-scale single-turn field coil (FC) locally applies a magnetic field to a sample, and a pickup loop (PL), which is concentric with the field coil and connected to flux-sensitive SQUID circuit, measures the sample's magnetic response~\cite{Gardner2001-gr, Huber2008-il, Kirtley2016-zz}. The SQUID susceptometer used in this study is gradiometric, with two counter-wound field coil-pickup loop pairs separated by $\sim$ 1 mm, only one of which (the ``front field coil'') is brought close to the sample surface [Figure~\ref{fig:squid}({\bf a}, {\bf b})]. Each FC-PL pair has a mutual inductance $|\Phi_\PL/I_\FC|$ of approximately $285\,\Phi_0/\mathrm{A}$, meaning that a current $I_\FC=1\,\mathrm{mA}$ flowing through the field coil threads a flux $|\Phi_\PL|=0.285\,\Phi_0\approx0.59\,\mathrm{mT}\cdot\um^2$ through the pickup loop. The gradiometric design of the device means that ideally in the presence of a sample with no magnetic response, e.g. a superconductor with London penetration depth $\lambda\to\infty$, the total mutual inductance between the two field coils and the SQUID is zero: $M_\infty = (\Phi^\mathrm{front}_\PL + \Phi^\mathrm{back}_\PL)/I_\FC \approx 0\,\Phi_0/\mathrm{A}$ (see Appendix~\ref{sec:data-processing}). The field coil, pickup loop, and shields are made of niobium (Nb).

When the front field coil is brought close to a superconducting sample, the sample screens the magnetic field from the field coil, reducing the mutual inductance of the front FC-PL pair, $\Phi^\mathrm{front}_\PL/I_\FC$, and thus the total mutual inductance of the susceptometer. The amount by which the sample modifies the SQUID mutual inductance is the scanning SQUID susceptibility signal,
\begin{equation}
    \begin{split}
        M(x, y)&=\left(\Phi^\mathrm{front}_\PL(x, y)+\Phi^\mathrm{back}_\PL\right)/I_\FC\\
        &=\Delta\Phi_\PL^\mathrm{front}(x, y)/I_\FC + M_\infty.
        \label{eq:def-susc}
        \end{split}
\end{equation}
$M$ is measured as a function of relative sample-sensor position $(x, y)$ as the sensor is raster-scanned over the sample surface at a fixed standoff distance $z_0$, or as a function of another parameter (e.g. temperature) for a fixed sensor position. $M$ is measured using low-frequency lock-in techniques to improve sensitivity and enable detection of both the in-phase magnetic response $M'(x, y)$ and out-of-phase magnetic response $M''(x, y)$ of the sample, which are related to $\chi'$ and $\chi''$, respectively. In such a measurement, the current through the field coil is
\begin{equation}
    \label{eq:I_FC}
    I_\FC(t)=I_{\FC,\,\pk}\cos(\omega t),
\end{equation}
where $I_{\FC,\,\pk}$ is the peak amplitude of the field coil current and $\omega$ is angular frequency of the lock-in amplifier excitation (typically $\omega/2\pi\sim 1\,\mathrm{kHz}$). The complex magnetic response, as measured by a lock-in amplifier, is then given by
\begin{equation}
    \label{eq:M-integral}
    M(x, y)=\frac{\sqrt{2}}{I_{\FC,\,\pk}}\int \Delta\Phi^\mathrm{front}_\PL(x, y, t)e^{-i\omega t}\,\mathrm{d}t,
\end{equation}
where the integral is taken over many AC cycles, as determined by the lock-in amplifier time constant. From this demodulated complex mutual inductance signal, we define $M'=\Re(M)$ and $M''=\Im(M)$. For the remainder of this work, we will refer to the front field coil simply as ``the field coil.''

If no vortices are present near the field coil, the in-phase component $M'$ is a direct measure of the local value of the sample's magnetic screening length (the London penetration depth $\lambda$ for bulk superconductors or $\Lambda=\lambda^2/d$ for thin films, where $d$ is the film thickness), which is in turn related to the superfluid density $n_s$~\cite{Kirtley2012-od}. Ideally, if there is no dissipation in the sample, $M$ is purely real. Dissipation and nonzero $M''$ can arise even far below the superconductor's critical temperature $\Tc$ due to motion of vortices under the Lorentz force (Eq.~\ref{eq:lorentz}) caused by the local applied AC field. This time-varying Lorentz force can cause the vortices to oscillate about the bottom of their pinning potentials or hop between pinning sites, leading to nonzero $M''$. This effect has been used to study anisotropic pinning in unconventional superconductors~\cite{Kalisky2011-ns, Zhang2019-hl, Bishop-Van_Horn2019-ng, Iguchi2021-wm}. The motion of vortices subject to a spatially uniform applied AC magnetic field has been imaged locally using scanning Hall probe microscopy~\cite{Kramer2010-hu, Raes2012-bo}.

In contrast to Refs.~\cite{Kalisky2011-ns, Zhang2019-hl, Bishop-Van_Horn2019-ng, Iguchi2021-wm}, here we examine the case where there are no vortices pinned near the SQUID when cooling through $\Tc$, i.e., a zero-field cooling situation. In such a scenario, dissipation due to vortex motion can only occur if vortices are first induced in the superconductor by the local applied AC field. Given the magnitude of the magnetic field that can be applied by the SQUID field coil (up to a few $\mT$), this mechanism is possible only for systems with a coherence length  exceeding a few hundred nanometers, for example a 2D Josephson junction array~\cite{Bishop-Van_Horn2022-lr} or a thin film close to its critical temperature.

\section{Experiment \& Modeling}
\label{sec:experiment}

We measured a sputtered Nb film with thickness $d=200\,\nm$ using a SQUID susceptometer with the geometry shown in Figure~\ref{fig:squid}({\bf a}, {\bf b}). The film has a variety of lithographically patterned structures (holes, slots, etc.), along with large regions of continuous Nb. The film is thick compared to the zero-temperature London penetration depth and coherence length of niobium ($\lambda(0),\,\xi(0) \lesssim 100\,\mathrm{nm}$). However, both the London penetration depth $\lambda(T)$ and the coherence length $\xi(T)$ diverge at $\Tc$ so that, for temperatures sufficiently close $\Tc$, the film may be considered 2D. Near $\Tc$, the film is magnetically 2D in the sense that $\Lambda(T)=\lambda^2(T)/d$ is large compared to the film thickness $d$. This means that the supercurrent density can be assumed to be uniform in the $z$ (out-of-plane) direction, giving a thickness-integrated sheet supercurrent density $\vec{K}_s=d\vec{J}_s$. Similarly, when $\xi(T)$ is large compared to $d$, the superfluid density $n_s=|\Psi|^2$ can be assumed to be uniform in the $z$ direction.

\subsection{Magnetic response of the continuous film}

We measured the local AC susceptibility in a region of continuous Nb film, roughly 25 $\um$ away from the film edge or any patterned features in the film. The front FC-PL pair of the SQUID susceptometer was positioned a distance $z_0\approx 0.5\,\um$ from the sample surface. The drive frequency for all measurements was fixed at $\omega/2\pi=500$ Hz. Further experimental details are provided in Appendix~\ref{sec:data-processing}.

\begin{figure}
    \centering
    \includegraphics[width=\linewidth]{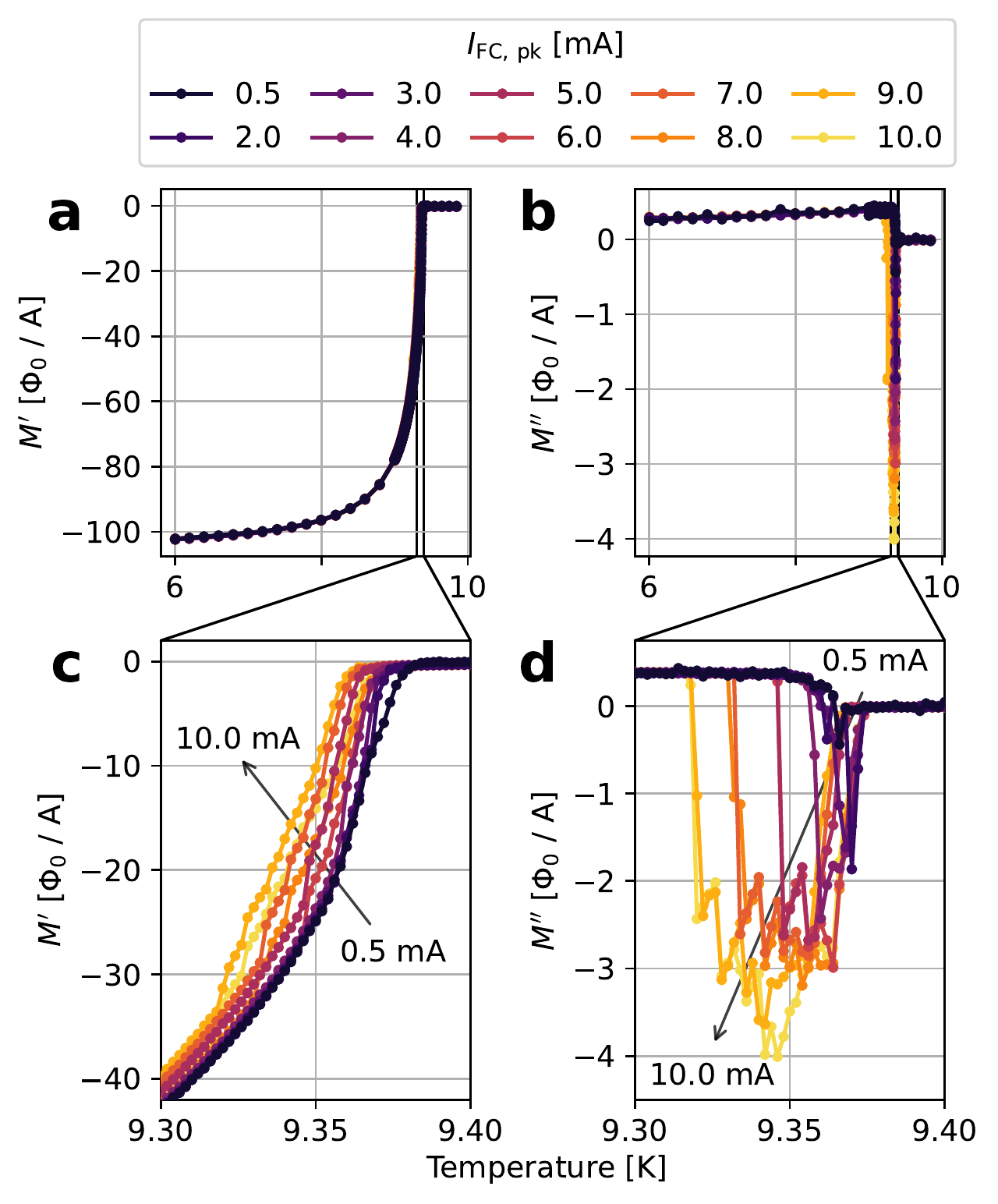}
    \caption{{\bf Measured temperature dependence of the complex magnetic response above a continuous Nb film.} ({\bf a}) In-phase and ({\bf b}) out-of-phase components of the complex magnetic response, $M=M'+iM''$, as a function of temperature for varying peak field coil currents. The second row shows ({\bf c}) $M'$ and ({\bf d}) $M''$ in the temperature range near $\Tc$ in which the variation of $M'$ and $M''$ can be attributed to vortex dynamics. Sharp steps in $M$ very close to $T_c$ in ({\bf c}, {\bf d}) are due to the dynamics of a small number of vortices induced in the film.}
    \label{fig:M-vs-T}
\end{figure}

Figure~\ref{fig:M-vs-T} shows the complex AC susceptibility, $M=M'+iM''$, as a function of temperature from 6 K to above $\Tc$ for peak field coil currents ranging from 0.5 mA to 10 mA, where $I_{\FC,\,\pk} = 10\,\mA$ corresponds to a maximum applied magnetic field of approximately 1.75 mT [Figure~\ref{fig:squid}({\bf c})]. At the lowest temperatures, $\lambda(T)$ is close to its zero-temperature value, $\lambda(0)$, and the film strongly screens the magnetic field from the field coil, reducing the mutual inductance between the field coil and pickup loop by approximately $(100\,\Phi_0/\mathrm{A})/(285\,\Phi_0/\mathrm{A}) \approx 35\%$. At these temperatures, $M$ is dominated by the superfluid response $M'$, the dissipative response $M''$ is small and roughly temperature-independent, and both components are independent of the applied field $H_\mathrm{applied}\propto I_{\FC,\,\pk}$, indicating a linear magnetic response. As $T$ approaches $\Tc$, $\lambda(T)$ diverges and the magnitude of superfluid response decreases rapidly as the film loses the ability to screen the applied field. As highlighted in Figure~\ref{fig:M-vs-T}({\bf c}, {\bf d}), at temperatures above 9.3 K, there is an $I_{\FC,\,\pk}$-dependent change in the slope of $M'(T)$, accompanied by a significant increase in the magnitude of the out-of-phase component $M''$, which is indicative of dissipation due to vortex dynamics induced by the applied AC field.

\begin{figure}
    \centering
    \includegraphics[width=\linewidth]{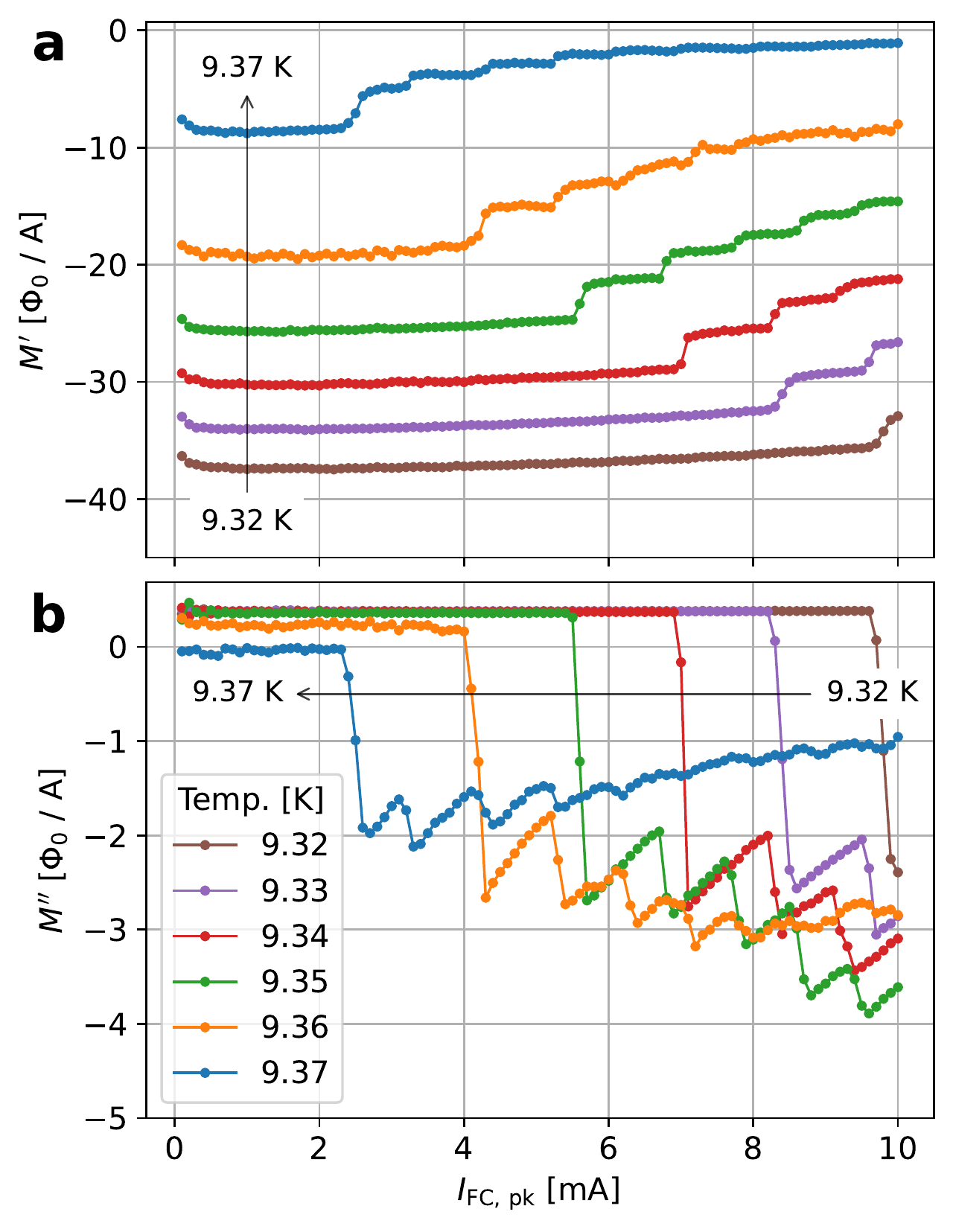}
    \caption{{\bf Measured fingerprints of few-vortex dynamics in the nonlinear magnetic response as a function of applied local AC magnetic field above a continuous Nb film.} The complex magnetic response $M=M'+iM''$, decomposed into ({\bf a}) its real part and ({\bf b}) its imaginary part. The distinct steps in both $M'$ and $M''$ with increasing field coil current are indicative of the dynamics of a small number of vortices induced in the film. The slight upturn in $M'$ at the lowest field coil currents $I_{\FC,\,\pk}$ is likely an artifact of the SQUID readout instrumentation. The weak increase in $M''$ with increasing temperature at small $I_{\FC,\,\pk}$ is likely due to a slight change in the temperature of the SQUID susceptometer, which is kept far below its critical temperature during the measurement (see Appendix~\ref{sec:data-processing}).}
    \label{fig:M-vs-Ifc}
\end{figure}

To further investigate the nonlinear, dissipative magnetic response observed near $\Tc$, we measured $M$ while varying the amplitude of the AC current through the field coil over the range $I_{\FC,\,\pk} = 0.1 - 10\,\mA$. Figure~\ref{fig:M-vs-Ifc} shows the complex magnetic response $M=M'+iM''$ as a function of $I_{\FC,\,\pk}$ for a series of temperatures near $T_c$.  At all temperatures, $M'$ starts out as approximately constant as $I_{\FC,\,\pk}$ is increased from zero, indicating that the local susceptibility is dominated by the linear superfluid response. As $I_{\FC,\,\pk}$ is increased further, $M'$ approaches zero in discrete steps, accompanied by a sawtooth pattern in the out-of-phase component of the signal, $M''$. This discrete, step-like pattern in the $M$ vs. $I_{\FC,\,\pk}$ curve results from the dynamics of a small number of vortex-antivortex pairs induced in the film by the SQUID susceptometry measurement. The value of $I_{\FC,\,\pk}$ at which the first step in $M$ occurs corresponds to the effective lower critical field of the film, which decreases rapidly as the temperature approaches $T_c$.

\subsection{Time-dependent Ginzburg-Landau modeling}
\label{sec:modeling}

The observed nonlinear behavior [Figure~\ref{fig:M-vs-Ifc}] is reminiscent of the results of measurements and TDGL simulations of the DC magnetization~\cite{Kanda2004-lx, Hernandez2002-nx}, and microwave magnetic response~\cite{Hernandez2002-oe, Hernandez2008-mi} of mesoscopic type-II superconductors subject to a spatially uniform applied magnetic field, where a step-like magnetic response arises primarily from the surface energy barrier. Our measurements differ from the scenarios considered in Refs.~\cite{Kanda2004-lx, Hernandez2002-nx, Hernandez2002-oe, Hernandez2008-mi} in two important respects. First, the applied magnetic field is not spatially uniform, rather it is applied by an asymmetric micron-scale current loop. Second, in contrast to Refs.~\cite{Hernandez2002-oe, Hernandez2008-mi}, the measurement timescale $2\pi/\omega=1/(500\,\mathrm{Hz})=2\,\mathrm{ms}$ is many orders of magnitude longer than the relevant relaxation time in the superconductor, as estimated below.

\begin{figure*}
    \centering
    \includegraphics[width=\linewidth]{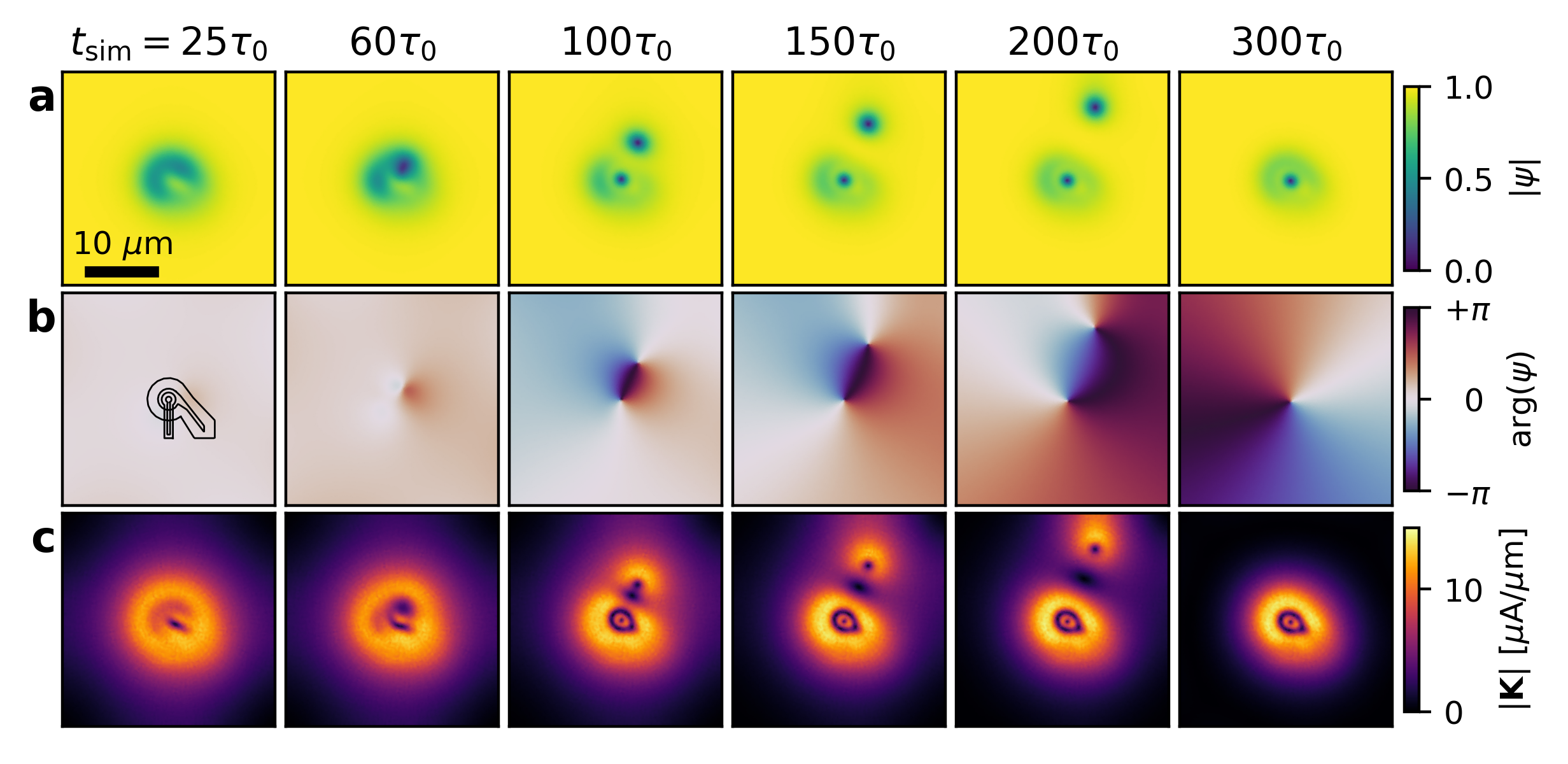}
    \caption{{\bf Generating and trapping a single vortex with the SQUID field coil.} Each column shows a snapshot of the output of a TDGL simulation taken at the time indicated at the top of the column. The simulation parameters are $\xi=0.9\,\um$, $\lambda=1.35\,\um$, $z_0=0.5\,\um$, and $I_\FC=2.5\,\mathrm{mA}$, where $I_\FC$ is the DC current flowing counterclockwise through the field coil. The initial state of the system at time $t_\mathrm{sim}=0$ is $\psi(\mathbf{r}, t_\mathrm{sim})=1$. From top to bottom, the rows show ({\bf a}) the magnitude of the order parameter $|\psi|$, ({\bf b}) the phase of the order parameter $\arg(\psi)$, and ({\bf c}) the magnitude of the sheet current density $|\mathbf{K}|$.
    Around $t_\mathrm{sim}=60\tau_0$ (second column), a vortex-antivortex pair is induced in the film and subsequently pulled apart by the Lorentz force (Eq.~\ref{eq:lorentz}) until the antivortex reaches the edge of the film around $t_\mathrm{sim}=250\tau_0$, leaving an isolated vortex trapped beneath the field coil. For $t_\mathrm{sim}>300\tau_0$, the system is stable indefinitely, with the supercurrent circulating counterclockwise just outside the vortex core and clockwise everywhere else. The SQUID field coil and pickup loop are drawn to scale in the first column of ({\bf b}).}
    \label{fig:squid-vortex}
\end{figure*}

To interpret the results shown in Figure~\ref{fig:M-vs-Ifc}, we have performed TDGL simulations (see Appendix~\ref{sec:tdgl} and Ref.~\cite{Bishop-Van_Horn2023-wr}) that take into account a realistic model of the SQUID sensor geometry [Figure~\ref{fig:squid}] and the dynamic, but low-frequency, nature of the AC susceptibility measurement. The characteristic time scale for this TDGL model is $\tau_0=\mu_0\sigma\lambda^2$, where $\mu_0$ is the vacuum permeability, $\sigma$ is the normal state conductivity of the film, and $\lambda$ is the London penetration depth (see Appendix~\ref{sec:tdgl}). Even assuming a very high normal state conductivity, e.g. $\sigma=(10^{-2}\mu\Omega\cdot\mathrm{cm})^{-1}=10^4\,\mathrm{S}/\um$~\cite{Webb1969-mb}, and noting that $\lambda(T)$ diverges at $\Tc$, this time scale is $\tau_0<\mu_0\cdot10^4\,\mathrm{S}/\um\cdot(2\,\um)^2\approx50\,\mathrm{ns}$ in the temperature range of our measurements, which is more than four orders of magnitude shorter than the measurement time scale $2\pi/\omega=2\,\mathrm{ms}$.

Thus, we will assume that in the absence of pinning, the order parameter $\Psi$ would be at equilibrium with the applied vector potential at all points in time. This ``quasistatic approximation'' has the following implications. (1) Vortices are allowed to reach a steady state configuration before the applied AC field changes significantly~\cite{Lemberger2016-pb}. (2) Only pinning-related losses will be captured by the model (cf. Section~\ref{sec:ac-loss}). (3) The simulated dynamics will be independent of the drive frequency $\omega$. The measurements support this assumption, as we see no qualitative difference in the nonlinear magnetic response if we change the drive frequency from $\omega/2\pi=500\,\mathrm{Hz}$ to $\omega/2\pi=5\,\mathrm{kHz}$.

The inputs to the TDGL model are the applied magnetic vector potential $\mathbf{A}_\mathrm{applied}(\mathbf{r})$ due to the SQUID field coil for a given field coil current $I_\FC$ [Figure~\ref{fig:squid}({\bf d})] and the relevant parameters of the Nb film, $\xi$, $\lambda$, and $d$. The outputs of the TDGL model include the normalized complex order parameter $\psi(\mathbf{r}, t)=\Psi(\mathbf{r}, t)/|\Psi_0|$ and the sheet current density in the film $\mathbf{K}(\mathbf{r},t)=\mathbf{K}_s(\mathbf{r}, t)+\mathbf{K}_n(\mathbf{r}, t)$, where $\Psi_0$ is the zero-field value of the order parameter, $\mathbf{K}_s$ is the sheet supercurrent density, and $\mathbf{K}_n$ is the sheet normal current density [Figure~\ref{fig:squid-vortex}]. From $\mathbf{K}$, we can calculate the resulting magnetic flux through the SQUID pickup loop, $\Phi_\PL$ to obtain the mutual inductance signal, $M$.

Because the film is much bigger than the SQUID field coil, and too large to model in its entirety, we model the geometry of the film as a square centered at the position of the SQUID with side length $L_\mathrm{film}=30\,\um$, which is approximately 10 times the outer radius of the field coil. In solving the TDGL model, we neglect the contribution of the induced currents in the film to the total magnetic vector potential in the film, i.e., we assume that the vector potential in the film is equal to $\mathbf{A}_\mathrm{applied}(\mathbf{r})$. This ``weak screening'' assumption is discussed further in Appendix~\ref{sec:tdgl}. These two approximations (modeling the film as a square with side length $L_\mathrm{film}$ and neglecting the induced vector potential) are likely to be the most significant sources of error in the modeling.

An example of a TDGL simulation with a static counterclockwise-flowing field coil current is shown in Figure~\ref{fig:squid-vortex}. Before any vortices are induced, the Meissner supercurrent flows clockwise and the normalized superfluid density $|\psi|^2$ is suppressed below the field coil [Figure~\ref{fig:squid-vortex}({\bf a}), left column], with the largest suppression occurring where the applied magnetic vector potential is largest. In Figure~\ref{fig:squid-vortex}, the peak applied field is just above the effective lower critical field of the film, $B_{c1}^\mathrm{eff}$, so a single vortex-antivortex pair is induced in the film near the edge of the field coil. Due to the Lorentz force (Eq.~\ref{eq:lorentz}), the vortex is pulled toward the center of the field coil and the antivortex is pushed away from the center of the field coil until it eventually exits the modeled domain, leaving a single isolated vortex trapped beneath the field coil.

\begin{figure*}
    \centering
    \includegraphics[width=\linewidth]{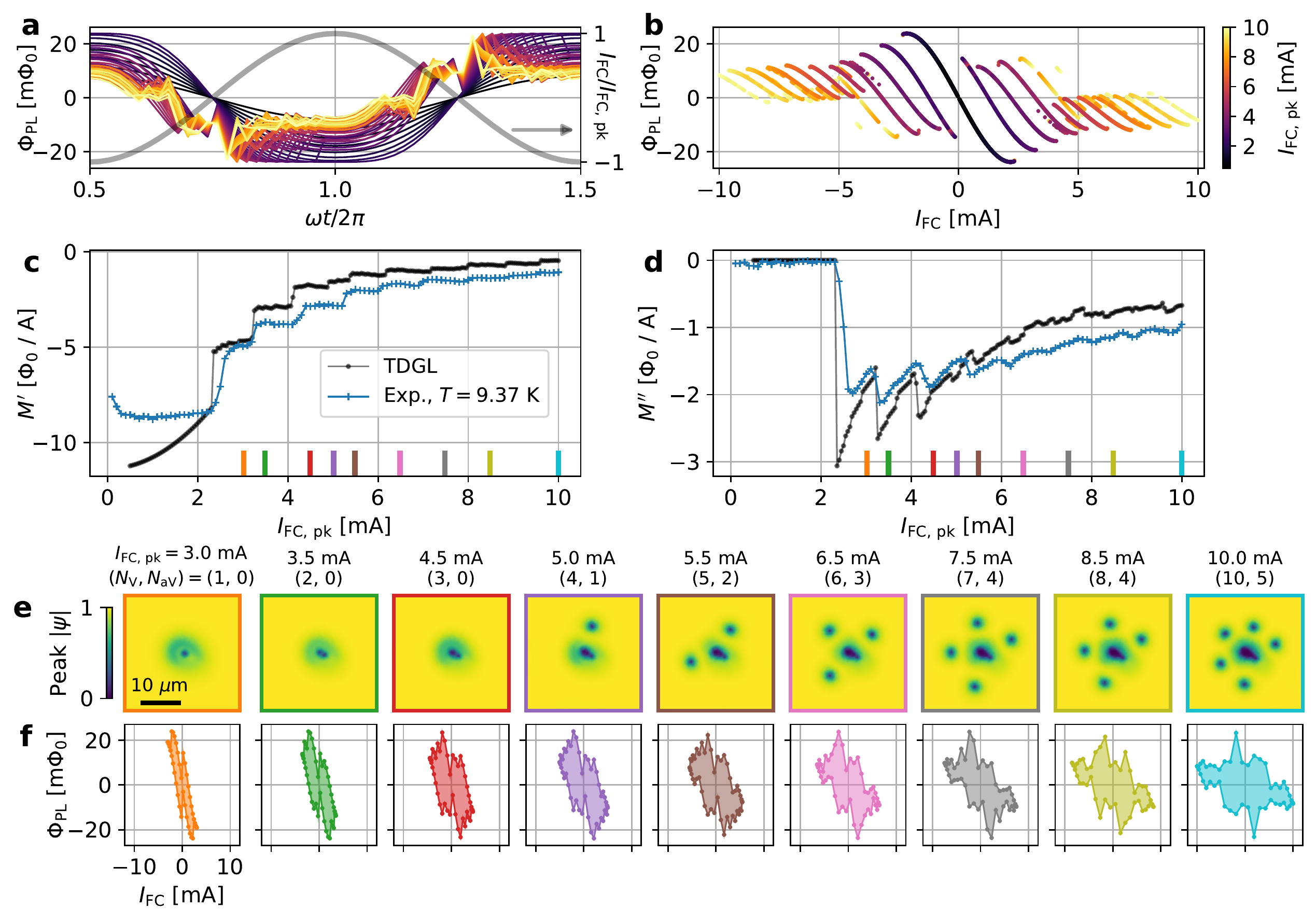}
    \caption{{\bf Simulated complex magnetic response as a function of applied AC field strength.} Simulated time-dependent flux through the SQUID pickup loop $\Phi_\PL(\omega t)$ due to Meissner currents in the Nb film over one AC cycle, plotted as a function of ({\bf a}) dimensionless time $\omega t$ and ({\bf b}) SQUID field coil current $I_\FC(\omega t)=I_{\FC,\,\pk}\cos(\omega t)$. To allow any hysteresis to accumulate, data from the first half cycle ($0 \leq \omega t < \pi$) is discarded. The normalized field coil current $I_\FC(\omega t)/I_{\FC,\,\pk}$ is shown in gray in ({\bf a}). For clarity, we show only every fifth value of $I_{\FC,\,\pk}$ in ({\bf a}). The branches in ({\bf b}) correspond to different configurations of vortices near the SQUID field coil. For a given value of $I_{\FC,\,\pk}$, the system transitions between branches throughout the AC cycle, tracing out a hysteresis loop as shown in ({\bf f}). ({\bf c}) Real part and ({\bf d}) imaginary part of the complex magnetic response $M=M'+iM''$ as a function of $I_{\FC,\,\pk}$. The measurement at $T=9.37$ K is shown in blue and the simulation results are shown in black. ({\bf e}) Magnitude of the order parameter $|\psi(\mathbf{r})|$ at the peak of the AC cycle for selected values of the peak field coil current $I_{\FC,\,\pk}$, indicated by the colored bars in ({\bf c}, {\bf d}). ({\bf f}) Hysteresis loops, $\Phi_\PL$ vs. $I_{\FC}$, for the selected values of $I_{\FC,\,\pk}$ indicated by the colored bars in ({\bf c}, {\bf d}). As in Figure~\ref{fig:squid-vortex}, the simulation parameters are $\xi=0.9\,\um$, $\lambda=1.35\,\um$, and $z_0=0.5\,\um$.}
    \label{fig:M-vs-Ifc-sim}
\end{figure*}

To simulate the low-frequency dynamics of the SQUID susceptometry measurement for a given peak field coil current $I_{\FC,\,\pk}$, we sample the AC field coil current (Eq.~\ref{eq:I_FC}) at a set of dimensionless times $\omega t_i$. At $\omega t_0=0$, we solve the TDGL model with the vector potential due to a field coil current $I_\FC(0)=I_{\FC,\,\pk}$, and with the initial condition $\psi(\mathbf{r}, 0)=1$, as in Figure~\ref{fig:squid-vortex}. Based on our quasistatic approximation, we allow the system to evolve until a steady-state is reached, then record the flux through the SQUID pickup loop, $\Phi_\PL(\omega t_i)$, due to the sheet current density $\mathbf{K}$ in the film. For all subsequent times, $\omega t_i$ with $i>0$,
we set the initial state of the TDGL simulation to be the final state found in the previous time step, $\omega t_{i-1}$, then allow the system to evolve until a new steady-state is reached. The complex SQUID susceptibility signal $M=M'+iM''$ is then calculated using Eq.~\ref{eq:M-integral}. To allow any hysteresis to accumulate, we simulate one and a half AC cycles and calculate $M$ using only $\Phi_\PL$ from the last full cycle.

The results of a TDGL simulation of the complex SQUID susceptibility signal $M$ are shown in Figure~\ref{fig:M-vs-Ifc-sim}. The top row shows the time-dependent flux through the pickup loop $\Phi_\PL(\omega t)$ due to the sheet current density in the film as a function of time [Figure~\ref{fig:M-vs-Ifc-sim}({\bf a})] and as a function of the instantaneous field coil current $I_\FC(\omega t)=I_{\FC,\,\pk}\cos(\omega t)$ [Figure~\ref{fig:M-vs-Ifc-sim}({\bf b})]. At the smallest values of the peak field coil current $I_{\FC,\,\pk}$, the magnetic response of the film is linear and non-hysteretic, so the flux through the  pickup loop due to the supercurrent in the film is $\Phi_\PL(\omega t)\propto -I_\FC(\omega t)$. As $I_{\FC,\,\pk}$ is increased, the film's response becomes nonlinear as the superfluid density is suppressed beneath the field coil, resulting in a ``squashed'' sinusoidal shape for $\Phi_\PL(\omega t)$. Although the magnetic response is nonlinear, it remains non-hysteretic until the peak applied field reaches the effective lower critical field of the film $B_{c1}^\mathrm{eff}$ around $I_{\FC,\,\pk}\approx 2.3\,\mathrm{mA}$, at which point a single vortex-antivortex pair is induced in the film during each half of the AC cycle.

Figure~\ref{fig:M-vs-Ifc-sim}({\bf b}) shows the flux through the pickup loop as a function of the instantaneous field coil current over one AC cycle, which is analogous to a traditional magnetization-field ($M-H$) curve~\cite{Clem1994-qu}. The plot consists of many branches, corresponding to different configurations of vortices $N=(N_\mathrm{V}, N_\mathrm{aV})$ near the SQUID field coil, where $N_\mathrm{V}$ is the number of vortices and $N_\mathrm{aV}$ is the number of antivortices. The central $N=(0, 0)$ branch intersects the origin, and the slope of this branch at the origin, $\left.\partial\Phi_\PL/\partial I_\FC\right|_{I_\FC=0}$, is the quantity measured in a linear SQUID susceptometry measurement.

Figure~\ref{fig:M-vs-Ifc-sim}({\bf c}, {\bf d}) shows the simulated complex magnetic response $M=M'+iM''$ as a function of $I_{\FC,\,\pk}$ (black curves) and the corresponding measurement at $T=9.37$ K (blue curves). Each step in $M$ as a function of $I_{\FC,\,\pk}$ corresponds to a different time-dependent vortex configuration $N$, driven by a competition between the vortex-(anti)vortex interaction and the vortex-field coil interaction that tends to pull vortices toward the center of the field coil and push antivortices away from the center of the field coil. The magnitude of the order parameter $|\psi(\mathbf{r})|$ at the peak of the AC cycle for selected values of $I_{\FC,\,\pk}$ is shown in Figure~\ref{fig:M-vs-Ifc-sim}({\bf e}).

As $I_{\FC,\,\pk}$ is increased from zero, the first three transitions are $N=(0, 0)\to(1, 0)\to(2,0)\to(3,0)$, where $N_\mathrm{aV}$ remains zero as the antivortices are pushed far away from the field coil and exit the film. At larger applied fields, the number of vortices beneath the field coil $N_\mathrm{V}$ increases, creating a potential well that traps a ring of $0<N_\mathrm{aV}<N_\mathrm{V}$ antivortices just outside the footprint of the field coil. Antivortices that are trapped in this potential well near the peak of the AC cycle [columns 4-9 of Figure~\ref{fig:M-vs-Ifc-sim}({\bf e})] annihilate with vortices trapped beneath the field coil when the applied field is reduced later in the AC cycle. Note that the steps in $M$ do not necessarily correspond to changes in the net number of vortices $(N_\mathrm{V}-N_\mathrm{aV})$ near the SQUID at the peak of the AC cycle.

Figure~\ref{fig:M-vs-Ifc-sim}({\bf f}) shows hysteresis loops, $\Phi_\PL$ vs. $I_\FC$, for the same values of $I_{\FC,\,\pk}$ shown in Figure~\ref{fig:M-vs-Ifc-sim}({\bf e}). Throughout the AC cycle for a given $I_{\FC,\,\pk}$, the system transitions between branches in Figure~\ref{fig:M-vs-Ifc-sim}({\bf b}), tracing out a hysteresis loop whose area is related to the energy dissipated in the film per cycle (i.e., the work done on the film by the lock-in amplifier)~\cite{Clem1991-vq}. Loops that stay on the central $N=(0, 0)$ branch in Figure~\ref{fig:M-vs-Ifc-sim}({\bf b}) have zero area and are dissipationless.

The model does not explicitly include any pinning centers (e.g. small defects in the film). However, within the framework outlined in Section~\ref{sec:ac-loss}, a vortex leaving the film as shown in Figure~\ref{fig:squid-vortex} is essentially a form of pinning. In the measurement, once the antivortex is pushed far from the field coil, it could leave the film as in the simulation (surface pinning), or become strongly pinned on a defect or annihilate with a vortex pinned elsewhere in the film (bulk pinning). In any of these cases, the antivortex is ``lost'' from the measurement, as it will never return to the vicinity of the SQUID or annihilate with its corresponding vortex.

Once the antivortex is lost due to any of these pinning scenarios, the vortex is trapped below the field coil by the supercurrent screening the applied field . It will remain trapped (metastably) even when the applied field is reduced below $B_{c1}^\mathrm{eff}$, leading to hysteresis in the film's AC magnetic response. This process, whereby one half of an induced vortex-antivortex pair is effectively lost from the measurement, is the origin of the step-like nonlinear magnetic response shown in Figure~\ref{fig:M-vs-Ifc}.

\subsection{Magnetic response near lithographically defined defects}
\label{sec:slot}

\begin{figure*}
    \centering
    \includegraphics[width=\linewidth]{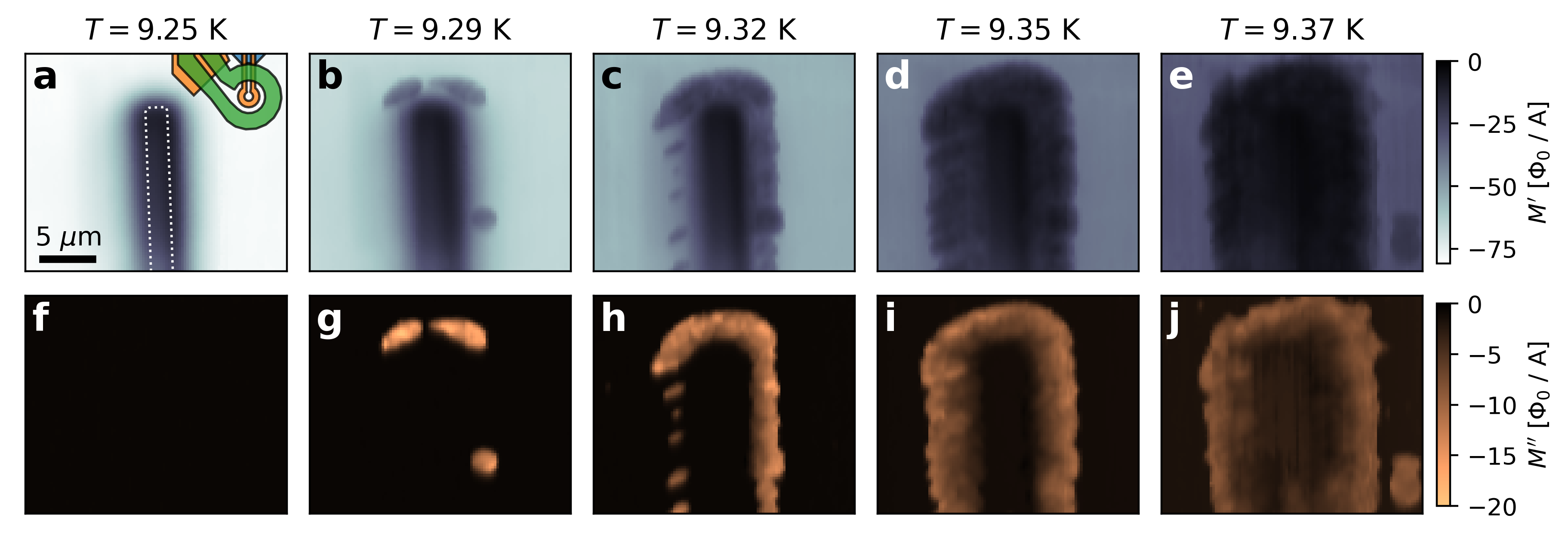}
    \caption{{\bf Measured spatial maps of vortex dissipation near a lithographically defined slot.} ({\bf a}-{\bf e}) Scans of the measured in-phase and ({\bf f}-{\bf j}) out-of-phase components of the complex AC susceptibility $M = M'+iM''$ near a $2\,\um$ wide slot at a series of temperatures close to $T_c$. At the lowest temperature ($T=9.25$ K), the AC susceptibility outside the slot is dominated by the superfluid response indicated by the strongly negative $M'$ ({\bf a}) and the nearly zero dissipative component $M''$ ({\bf f}). At higher temperatures, inhomogeneity in $M'$ appears around the slot accompanied by non-zero $M''$ due to vortex dynamics induced by the local applied AC field. The SQUID susceptometer is drawn to scale in ({\bf a}), and the scale bar in ({\bf a}) applies to all panels. The dotted white line in ({\bf a}) shows the dimensions and approximate location of the slot.}
    \label{fig:SUSC_Tseries_scans}
\end{figure*}

We also imaged the local AC susceptibility near a lithographically defined defect in the film, a $2\,\um$ wide slot, by raster scanning the SQUID in a plane parallel to and a few hundred nm above the Nb film. The drive frequency of the field coil was again fixed at $\omega/2\pi=500$ Hz with an amplitude of $I_{\FC,\,\pk} = 4\,\mA$.

Figure \ref{fig:SUSC_Tseries_scans} shows susceptibility scans at a series of temperatures near $T_c$. At the lowest temperature, the signal is dominated by the superfluid response, as indicated by a strongly negative and spatially uniform $M'$ outside the slot
[Figure~\ref{fig:SUSC_Tseries_scans}({\bf a})] and nearly zero dissipative ($M''$) response [Figure~\ref{fig:SUSC_Tseries_scans}({\bf f})]. As the temperature is increased toward $\Tc$, the superfluid response decreases toward zero [Figure~\ref{fig:SUSC_Tseries_scans}({\bf b}-{\bf e})] while a non-zero dissipative response nucleates near the edge of the slot and expands outward with increasing temperature [Figure~\ref{fig:SUSC_Tseries_scans}({\bf g}-{\bf j})].

\begin{figure*}
    \centering
    \includegraphics[width=\linewidth]{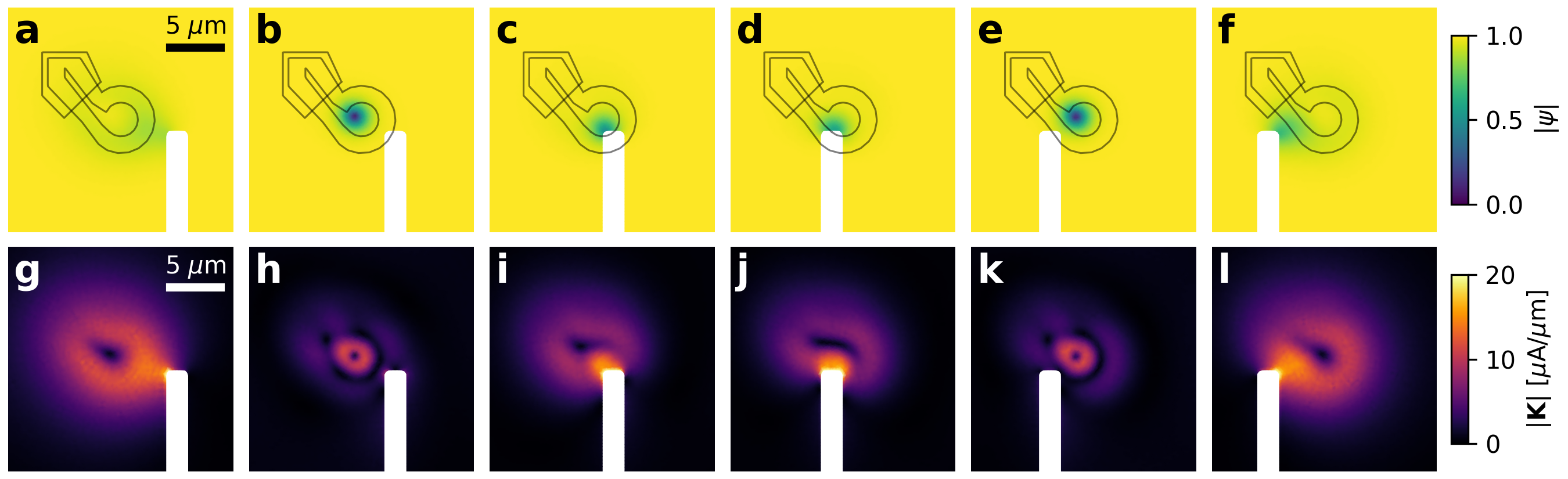}
    \caption{{\bf Simulation of vortex nucleation near a slot.} The steady-state values of the magnitude of the order parameter $|\psi|$ ({\bf a}-{\bf f}) and the sheet current density $|\mathbf{K}|$ ({\bf g}-{\bf l}) are shown as a function of the relative position between the slot and the SQUID susceptometer. For this DC simulation, the parameters are $I_\mathrm{FC}=1.1\,\mA$, $\xi=0.9\,\um$, $\lambda=1.35\,\um$, and $z_0=0.5\,\um$. For this value of $I_\mathrm{FC}$, as the SQUID is scanned over the top edge of the slot, vortices are induced when the SQUID is near the corners of the slot, but not when the SQUID is near the center of the top edge. This result is consistent with Figure~\ref{fig:SUSC_Tseries_scans}({\bf b}, {\bf g}), where the dissipative signal first appears near the corners of the slot. As in Figures~\ref{fig:squid-vortex} and \ref{fig:M-vs-Ifc-sim}, the modeled domain is a square with side length $L_\mathrm{film}=30\,\um$ centered at the center of the field coil, but for clarity only the central $20\,\um\times 20\,\um$ are shown. The slot is modeled with a width of $2\,\um$ and corners rounded with a radius of $0.1\,\um$. Each column corresponds to a different slot position relative to the SQUID.}
    \label{fig:squid-slot-top}
\end{figure*}

\begin{figure*}
    \centering
    \includegraphics[width=\linewidth]{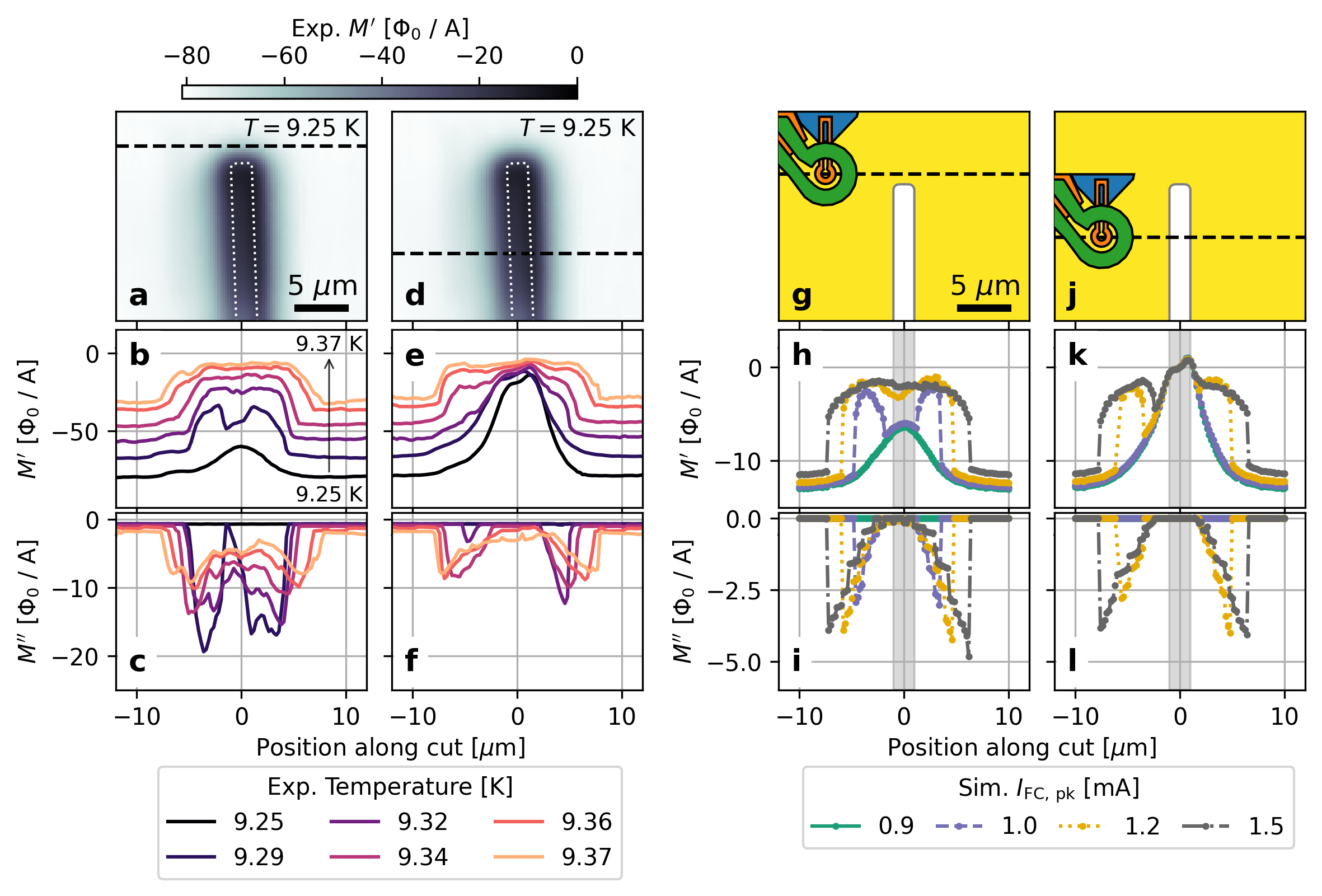}
    \caption{{\bf Measured (a-f) and simulated (g-l) vortex dynamics near a lithographically defined defect.} ({\bf a}, {\bf d}) Measured in-phase magnetic response $M'$ near the slot at $T=9.25$ K, with $I_{\FC,\,\pk}=4$ mA. The dashed black line in ({\bf a}) indicates the location of line cut near the top of the slot for which $M'$  and $M''$ are shown as a function of temperature in ({\bf b}) and ({\bf c}), respectively. The dashed black line in ({\bf d}) indicates the location of the line cut several microns below the top of the slot for which $M'$  and $M''$ are shown as a function of temperature in ({\bf e}) and ({\bf f}), respectively. ({\bf g} - {\bf l}) Simulated complex magnetic response as a function the relative position between the SQUID and the slot, which are drawn to scale in ({\bf g}, {\bf j}). The dashed black line in ({\bf g}) indicates the location of the line cut near the top of the slot for which $M'$ and $M''$ are shown as a function of $I_{\FC,\,\pk}$ in ({\bf h}) and ({\bf i}), respectively. The dashed black line in ({\bf j}) indicates the location of the line cut several microns below the top of the slot for which $M'$ and $M''$ are shown as a function of $I_{\FC,\,\pk}$ in ({\bf k}) and ({\bf l}), respectively. The parameters of the TDGL simulation are $\xi=0.9\,\um$, $\lambda=1.35\,\um$, and $z_0=0.5\,\um$.}
    \label{fig:squid-slot-cuts}
\end{figure*}

The distinctive spatial dependence of $M$ in Figure~\ref{fig:SUSC_Tseries_scans} can be understood by considering that the surface energy barrier is lower than the barrier to generate a vortex-antivortex pair in the bulk. Thus, for a given field coil current, dissipation due to induced vortex dynamics occurs at a lower temperature when the field coil is near the edge of the slot. Given the rapid decrease in $\xi(T)$ with decreasing temperature near $\Tc$ and the technical requirement that granularity of the finite element mesh used to solve the TDGL model be small compared to $\xi(T)$, it is not computationally practical to use the methods described above to model the temperature dependence of $M$ as a function of position near the slot. Furthermore, treating the film as 2D and neglecting the induced vector potential both become worse approximations at lower temperature. Instead, to qualitatively investigate the spatial dependence of the magnetic response, we model the system at a fixed temperature (i.e., fixed values of $\xi$ and $\lambda$) and vary $I_{\FC,\,\pk}$ as a proxy for varying the temperature.

Figure~\ref{fig:squid-slot-top} shows a TDGL simulation of the steady-state response of a superconducting film (with the same material parameters as in Figures~\ref{fig:squid-vortex} and \ref{fig:M-vs-Ifc-sim}) to a DC current of $I_\mathrm{FC}=1.1\,\mA$ in the field coil as the SQUID is scanned across the top of a $2\,\um$ wide slot in the film. Although this value of $I_{\FC}$ is less than half the value of $I_{\FC,\,\pk}$ at which vortex dynamics begin to occur far from the film edge [Figure~\ref{fig:M-vs-Ifc-sim}], a vortex is induced in the film when the field coil is near the corners of the slot, where the surface energy barrier is reduced due to geometrical current crowding~\cite{Clem2011-ji}. In contrast to Figures~\ref{fig:squid-vortex} and \ref{fig:M-vs-Ifc-sim}, the vortices in Figure~\ref{fig:squid-slot-top} nucleate at the superconductor-insulator interface at the boundary of the slot and do not require a corresponding antivortex to be induced in the film. This result is consistent with the measurement shown in Figure~\ref{fig:SUSC_Tseries_scans}, where at $T=9.29$ K a dissipative magnetic response first appears when the SQUID is near the corners of the slot [Figure~\ref{fig:SUSC_Tseries_scans}({\bf b}, {\bf g})].

Figure~\ref{fig:squid-slot-cuts} shows line cuts of the measured $M(T)$ [Figure~\ref{fig:squid-slot-cuts}({\bf a}-{\bf f})] and simulated $M(I_{\FC,\,\pk})$ [Figure~\ref{fig:squid-slot-cuts}({\bf g}-{\bf l})] across the slot. The simulations qualitatively capture the spatial dependence of the dissipationless magnetic response [$T=9.25$ K in ({\bf b}-{\bf f}) and $I_{\FC,\,\pk}=1.0$ mA in ({\bf h}-{\bf l})] and the hysteretic response due to vortex dynamics [$T>9.25$ K in ({\bf b}-{\bf f}) and $I_{\FC,\,\pk}>1.0$ mA in ({\bf h}-{\bf l})]. When the SQUID is near the top of the slot, the magnetic response is roughly symmetric about the center of the slot. Vortex dynamics first begin to occur near the corners of the slot, with the spatial extent of the dissipative response widening with both increasing temperature and increasing AC field strength. When the SQUID is below the top of the slot, the magnetic response is asymmetric about the center of the slot due to the geometry of the SQUID, and vortex dynamics begin to occur at a slightly higher temperature and higher AC field strength than when the SQUID is near the top of the slot. The measured $M(T)$ shows broadened features compared to the simulated $M(I_{\FC,\,\pk})$, likely due to thermal fluctuations or other stochastic effects. Although the comparison shown in Figure~\ref{fig:squid-slot-cuts} is only qualitative due to the technical limitations described above, measurements of $M(T)$ in mesoscopic thin film superconductors would be more amenable to more quantitative modeling, which could allow for self-consistent estimation of the temperature dependence of $\xi$, $\lambda$, and the vortex unbinding and surface free energy barriers.

\section{Discussion}
\label{sec:discussion}

In this section, we discuss our results as they relate to previous theoretical, computational, and experimental studies of local induced vortex dynamics in thin films. Lemberger, \textit{et al.} developed a theory for the lower critical field of a large thin film where the magnetic field is applied by a small coil or point dipole at the center of the film~\cite{Lemberger2013-lu, Lemberger2013-ha}. They calculated the applied magnetic field at which a state with a single vortex-antivortex pair in the film has a lower free energy than a vortex-free state (that is, the thermodynamic critical field for a dipole-like source, $B_{c}^\mathrm{d}$)~\cite{Lemberger2013-lu}. They also calculated the maximum applied field for which a metastable vortex-free state exists, $B_0^\mathrm{crit} > B_{c}^\mathrm{d}$, which is the applied field above which vortices must be present in the film~\cite{Lemberger2013-ha}. The thermodynamic critical field in the strong screening limit for a point dipole field source was found to be
\begin{equation}
    B_{c}^\mathrm{d}=\frac{\sqrt{2}}{\rho_0}\frac{\Phi_0}{2\pi\Lambda}\ln\left(\frac{2\Lambda}{\xi}\right),
    \label{eq:B0c1}
\end{equation}
where $\Lambda=\lambda^2/d$ and $\rho_0$ is the radial distance from the field source at which the applied field changes sign~\footnote{The strong and weak screening regimes are characterized by $\Lambda \ll \rho_0$ and $\Lambda \gg \rho_0$, respectively}. An approximation for $B_0^\mathrm{crit}$, again for a point dipole field source, which applies for both strong and weak screening, is given by
\begin{equation}
    B_0^\mathrm{crit}\approx\frac{\Phi_0}{2\pi\xi}\left(\frac{3\sqrt{6}}{\rho_0}+\frac{1}{2\Lambda}\right).
    \label{eq:B0crit}
\end{equation}
Due to the free energy barrier for generating a vortex-antivortex pair, $B_{c}^\mathrm{d}$ is much smaller than the applied field at which vortices actually begin to appear. These values therefore provide bounds on the applied field $B_{c}^\mathrm{d} < B_{c1}^\mathrm{eff} < B_0^\mathrm{crit}$ at which a vortex-antivortex pair will be induced in the film, with the actual value of $B_{c1}^\mathrm{eff}$ being determined by the size of the vortex-antivortex unbinding barrier relative to the thermal energy $k_\mathrm{B}T$. In the case of an infinite film, the lowest energy vortex-bearing state was found to consist of a vortex near the field source and an antivortex far (but not infinitely far) from the field source~\cite{Lemberger2013-lu}.

In our case, the magnetic field source is a small, asymmetric field coil rather than a point dipole, and the maximum value of the applied field for a given field coil current is $B_\mathrm{max}\approx 0.16\,\mT/\mA\times I_\FC$, and the applied field changes sign roughly at a distance $\rho_0\approx3\,\um$ from the center of the field coil. Using the same parameters as in the TDGL simulations above, $\lambda=1.35\,\um$ and $\xi=0.9\,\um$, we have $B_{c}^\mathrm{d}\approx0.05\,\mT$ and $B_0^\mathrm{crit}\approx0.9\,\mT$, whereas the first vortex-antivortex pair appears in our measurements and TDGL simulations at roughly an applied field of $0.16\,\mT/\mA\times 2.3\,\mA \approx 0.37\,\mT$. Thus, the bounds provided by Eqs.~\ref{eq:B0c1} and \ref{eq:B0crit} are rather loose.

The depairing critical current density $J_c$ is the maximum dissipationless current density that a superconductor can stably support within Ginzburg-Landau theory, under the assumption that the superfluid density $|\Psi|^2$ and supercurrent density $\mathbf{J}_s$ are spatially uniform~\cite{Tinkham2004-ln}. In SI units, $J_c$ is given by
\begin{equation}
    J_c=\left(\frac{2}{3}\right)^{3/2}\frac{B_c}{\mu_0\lambda}=\frac{1}{3\sqrt{3}}\frac{\Phi_0}{\mu_0\pi\xi\lambda^2},
    \label{eq:Jc-GL}
\end{equation}
where $B_c=\Phi_0/(2\sqrt{2}\pi\xi\lambda)$ is the thermodynamic critical field assuming a uniform applied field. For the parameters used in our TDGL simulation, this yields a depairing sheet current density of $J_cd\approx 12.3\,\uA/\um$. For the largest vortex-free DC field coil current in our simulation, $I_{\FC,\,\pk}=2.265\,\mA$, the maximum sheet current density in the film was $|\mathbf{K}|_\mathrm{max}=14.1\,\uA/\um$. There are two reasons that the simulated $|\mathbf{K}|_\mathrm{max}$ may exceed the depairing sheet current density $J_cd$. First, in the simulation we have neglected the induced vector potential due to Meissner currents in the film. For a given value of the applied vector potential, this approximation will lead to a slightly different value of the Meissner supercurrent density than a fully self-consistent solution that takes the induced vector potential into account. Re-running the simulation at $I_{\FC,\,\pk}=2.265\,\mA$ with the induced vector potential included (which is computationally much more costly), we find $|\mathbf{K}|_\mathrm{max}=14.3\,\uA/\um$, indicating that neglecting screening is not a large source of error in this case. Second, $J_c$ is derived assuming a spatially uniform superfluid density and supercurrent density~\cite{Tinkham2004-ln}, whereas both $|\Psi|^2$ and $\mathbf{J}_s$ are inhomogeneous for a local applied field just below the effective lower critical field. Nonetheless, our results suggest that the condition $|\mathbf{K}|_\mathrm{max}\approx J_cd$ provides a reasonable estimate for the maximum vortex-free sheet supercurrent density in thin films with a local magnetic field source, despite the highly inhomogeneous supercurrent distribution.

In transport measurements of thin film type-II superconductors, especially those that are narrow transverse to the direction of current flow such as superconducting nanowire single photon detectors (SNSPDs), it is thought that finite resistance due to vortex motion first occurs when the supercurrent density at the edge exceeds the depairing current density $J_c$, causing vortices to nucleate at the edge and then travel into the bulk~\cite{Vodolazov2003-ya}. This process is likely dominated by defects at the edge~\cite{Cerbu2013-hr, Benfenati2020-mg, Bezuglyj2022-cm} or geometrical current crowding~\cite{Clem2011-ji, Clem2012-og, Berdiyorov2012-rn, Hortensius2012-op, Henrich2012-jm, Adami2013-ci, Ilin2014-od, Jonsson2022-mb} (for example at sharp corners), both of which lead to an inhomogeneous supercurrent distribution at the film edge. The result is that the measured spatially-averaged critical current density $I_c/(wd)$ (where $I_c$ is the transport critical current, $w$ is the film width, and $d$ is the film thickness) is far below the depairing current density. The same considerations apply to magnetic measurements where the field applied at the edge is nonvanishing, as demonstrated in Figures~\ref{fig:squid-slot-top} and \ref{fig:squid-slot-cuts}. Measuring the point at which a locally applied field generates a vortex-antivortex pair far from the edge, as we have done here, can provide a more direct probe of the intrinsic depairing current density.

Our measurements are consistent with a scenario where, once antivortices have been generated, they can be pushed tens of microns away from the field coil without becoming pinned on defects within the sensing area of the SQUID (i.e., the area where the Meissner current due to the field coil is nonvanishing) before exiting the film or becoming pinned far from the SQUID. This is in contrast to an earlier computational study of vortex dynamics in AC susceptibility measurements~\cite{Lemberger2016-pb} in which a uniform density of strong pinning defects was assumed. SQUID susceptometry measurements of films with a higher density of pinning centers would likely show somewhat different nonlinear behavior, particularly if induced vortices had a large probability of becoming pinned within the sensing area of the SQUID. Gardner, \textit{et al.}~\cite{Gardner2002-un} generated vortex-antivortex pairs in a cuprate thin film using a DC current in the field coil of a scanning SQUID susceptometer. In Ref.~\cite{Gardner2002-un}, the vortex-antivortex pairs were pinned close to the field coil after being generated, and these pinned vortices were subsequently imaged directly using the scanning SQUID sensor.

Previous scanning SQUID susceptometry measurements of a niobium film near its critical temperature showed telegraph-like fluctuations of the complex AC susceptibility as a function of time, with the magnitude of the fluctuations being consistent with a single flux quantum appearing or disappearing near the sensor~\cite{Wissberg2018-lq}. Scanning SQUID measurements performed on NbTiN films near the thickness-tuned superconductor-to-insulator transition showed similar telegraph-like fluctuations near ``puddles'' where the superfluid density or critical temperature was locally suppressed~\cite{Kremen2018-dp}. While the field coil current amplitude dependence of the fluctuations was not discussed in Refs.~\cite{Wissberg2018-lq, Kremen2018-dp}, these previous measurements can be viewed as a stochastic version of the present work, where the peak of the applied AC field is below the effective lower critical field and the vortex unbinding energy barrier is overcome at random by thermal fluctuations. Other methods for controllably generating vortex-antivortex pairs in superconducting thin films include locally heating the film while driving a transport current~\cite{Ge2017-wa} and coupling the film to a ferromagnetic structure~\cite{Gladilin2009-zl, Simmendinger2020-pv}.

Trapped vortices are a known source of energy loss in superconducting qubits and resonators~\cite{Song2009-ps, Song2009-dg, Kroll2019-co, Eley2021-lu}. In many cases, the effects of vortex trapping due to ambient magnetic fields can be mitigated by magnetic shielding and/or designing devices to exclude or strongly pin vortices~\cite{Stan2004-ec, Kuit2008-gb, Song2009-ps}. Vortices can also be induced in superconducting circuits by pulsed control fields~\cite{Song2009-dg}, for example, those delivered by on-chip flux bias lines in frequency-tunable superconducting qubits. As feature sizes become smaller in the quest to miniaturize superconducting qubits, classical superconducting logic circuits, and related devices, induced vortices will likely become an even more important factor in device performance. The numerical methods described here can be used to optimize the design of features such as on-chip flux bias lines to mitigate vortex related effects. Understanding vortex nucleation and motion in mesoscopic superconductors subject to nonuniform magnetic fields is also likely to be critical in interpreting measurements of emerging superconductor-semiconductor-ferromagnetic insulator hybrid systems~\cite{Liu2020-hf, Vaitiekenas2022-to, Razmadze2023-rt}.

\section{Conclusion}
\label{sec:conclusion}

In this work, we have experimentally and computationally studied the dynamics of vortices in low-frequency AC susceptibility measurements of thin film superconductors where the field coil and pickup loop are of comparable size to the Ginzburg-Landau coherence length. The measurements reveal discrete steps in the complex magnetic response as a function of the strength of the applied AC field. Time-dependent Ginzburg-Landau simulations allow us to identify the dynamic behavior of a small number of induced vortex-antivortex pairs responsible for these steps.

In future work, it would be preferable to measure films that are two-dimensional at all temperatures (i.e., films with thickness $d<\xi(0),\lambda(0)$ to better satisfy the assumptions of the modeling), and to investigate more systematically the dependence of the induced vortex dynamics on the size/geometry of the superconductor and pinning centers. The TDGL model employed here (Appendix~\ref{sec:tdgl} and Ref.~\cite{Bishop-Van_Horn2023-wr}) allows for modeling pinning centers of various geometries in the form of holes or regions of reduced or vanishing $\Tc$, which could be engineered, for example, by ion irradiation~\cite{Massee2015-el, Aichner2019-zq} or by depositing superconducting islands on 2D normal metal or semiconducting substrates~\cite{Resnick1981-ow, Eley2011-id, Han2014-yv, Bottcher2018-fx, Naibert2021-ps, Bishop-Van_Horn2022-lr, Al_Luhaibi2022-cl}. Arrays of such islands are model systems for exploring the physics of the superconductor-to-insulator quantum phase transition~\cite{Eley2011-id, Han2014-yv, Bottcher2018-fx, Sacepe2020-eo}.

Studying local induced vortex dynamics at higher frequencies, where the quasistatic approximation breaks down, would open up two new avenues for investigation. First, if we neglect viscous losses, the energy dissipated per AC cycle is frequency-independent, but the energy dissipated per unit time (i.e., the average dissipated power) grows linearly with frequency. At some point, self-heating due to vortex motion will cause additional nonlinearity and thermal hysteresis in the system. Second, at high enough frequencies, viscous flux flow losses will become significant, making the vortex dynamics strongly frequency-dependent~\cite{Clem1991-vq}. Vortex motion at higher frequencies is more relevant to superconducting quantum and classical logic circuits, which typically operate in the GHz range. One could also induce vortices with a DC current in the field coil and probe their dynamics with a smaller superimposed AC excitation.

In summary, we have developed a predictive numerical model of low-frequency vortex dynamics induced by scanning SQUID susceptometry and applied the model to measurements of the nonlinear, dissipative magnetic response of a niobium film near its critical temperature.
Our results highlight the fact that the dynamics of vortices in AC susceptibility measurements of thin films can be sensitive to finite-size effects, even when the source of the applied field is much smaller than the film being measured. While finite-size effects can impact the dynamics of vortices after they have been generated, the process of inducing a vortex-antivortex pair with a locally applied field far from the edge is not sensitive to nonidealities of the edge. Our local approach is therefore complementary to global AC susceptibility and transport measurements, as it allows one to investigate and/or minimize the impact of edge defects and spatial inhomogeneities in studies of vortex-related dissipation in thin film superconductors. Local measurements also allow one to map out dissipation due to vortex dynamics near defects and interfaces. Insights provided by local magnetic response measurements and detailed numerical modeling are relevant to superconducting quantum circuits and sensors, and to studies of the breakdown of superconductivity in low-dimensional systems~\cite{Sacepe2020-eo}.

\section*{Author contributions}
E.M. performed the scanning SQUID measurements. L.B.V.H. performed the modeling and wrote the manuscript with input from all authors. K.A.M. supervised the project.

\section*{Data \& code availability}
The scanning SQUID data, simulation results, simulation code, and code used to generate the figures are available at Ref.~\footnote{\href{https://github.com/loganbvh/vortex-dynamics-induced-by-scanning-squid}{https://github.com/loganbvh/vortex-dynamics-induced-by-scanning-squid}, \href{https://doi.org/10.5281/zenodo.7857363}{https://doi.org/10.5281/zenodo.7857363}}.

\begin{acknowledgments}
This work is supported by the Department of Energy, Office of Basic Energy Sciences, Division of Materials Sciences and Engineering, under contract DE-AC02-76SF00515. Some of the computing for this project was performed on the Sherlock cluster at Stanford University. We would like to thank Stanford University and the Stanford Research Computing Center for providing computational resources and support that contributed to these research results. We would like to thank John R. Kirtley for providing feedback on this manuscript. We acknowledge E. Track, M. Stoutimore, and V. Talanov of Northrop Grumman Mission Systems for providing samples for this work.
\end{acknowledgments}

\appendix

\section{Experimental details}
\label{sec:data-processing}

The sample is attached to the top of a copper sample stage using cryogenic grease. A resistive heater and calibrated silicon diode temperature sensor are attached to the bottom side of the same copper stage to control the sample temperature. The niobium film is deposited on top of a 500 $\um$ Si/SiO${}_2$ substrate, and is therefore separated from the heater/thermometer by the copper sample stage, cryogenic grease, and thick insulating substrate, all of which introduce some thermal impedance. As a result, there may be a thermal gradient between the heater/thermometer and the niobium film, such that the temperature recorded from the thermometer is higher than the actual temperature of the film. In particular, the top surface of the sample may be cooled by residual helium-4 exchange gas in the sample volume, or by mechanical contact with the SQUID susceptometer, which is kept a few Kelvin colder than the sample throughout the measurement. The exact critical temperature of the film is not relevant to our analysis of the vortex dynamics.

For the data presented in Figures~\ref{fig:M-vs-T} and \ref{fig:M-vs-Ifc}, we have subtracted the sample-independent SQUID mutual inductance $M_\infty$ and numerically corrected the lock-in amplifier phase. In practice, $M_\infty$, which we refer to as the ``field coil imbalance,'' is nonzero due to minor lithographic imperfections in the SQUID chip. The typical magnitude of the field coil imbalance is a few percent of the mutual inductance of a single field coil-pickup loop pair.

The value of $M_\infty$ depends weakly on the temperature of the SQUID susceptometer. The susceptometer is thermally isolated from the sample so that the sensor remains superconducting even when the sample is heated above its critical temperature. However, the isolation is not perfect and the temperature of the sensor does increase slightly when heating the sample. This residual thermal coupling between the SQUID and the sample is likely the origin of the weak increase in $M''$ with increasing sample temperature in Figure~\ref{fig:M-vs-Ifc}({\bf b}).

To correct the lock-in amplifier phase starting with the raw complex mutual inductance $M_0=M_0'+iM_0''$, we fit $M_0''(T)$ vs. $M_0'(T)$ measured with $I_{\FC,\,\pk}=0.5\,\mA$ [black points in Figure~\ref{fig:M-vs-T}] for $9\,\mathrm{K}\leq T\leq 9.3\,\mathrm{K}$ to a linear model, $M_0''=aM_0' + b$, calculate the angle $\alpha=\arctan(a)$, and define a rotated mutual inductance according to $M_1=M_0e^{-i\alpha}$. This temperature range is chosen because there is a large change in $|M_0|$ over this range, but no significant nonlinearity or evidence of vortex dynamics. After rotation, the out-of-phase component $\Im(M_1)$ is consistent with Gaussian noise over this temperature range, indicating that the nonzero $\mathrm{Im}(M_0)$ was indeed due to an offset in the lock-in amplifier phase and not an actual dissipative response. We then calculate $M_\infty$ by taking the mean of $M_1(T)$ for $T\geq 9.4\,\mathrm{K}$, which is well above $\Tc$ where the sample has no magnetic response, and define $M_2=M_1-M_\infty=M_0e^{-i\alpha}-M_\infty$, which is the quantity plotted in the main text. The same values of $\alpha$ and $M_\infty$, found from the $I_{\FC,\,\pk}=0.5\,\mA$ dataset, are applied to all curves shown in Figures~\ref{fig:M-vs-T} and \ref{fig:M-vs-Ifc}, which were measured in the same cooldown.

The same process was applied for the data shown in Figures~\ref{fig:SUSC_Tseries_scans} and \ref{fig:squid-slot-cuts}, which were acquired in a different cooldown. The phase offset $\alpha=\arctan(a)$ was found by fitting the raw data measured at $T=9.25$ K pixel-wise to a linear model, $M_0''=aM_0' + b$, where again this dataset was chosen because there was no evidence of nonlinearity or dissipation [Figure~\ref{fig:SUSC_Tseries_scans}({\bf a})]. $M_\infty$ was found by taking the mean of $M$ over a scan measured at $T=9.7$ K, well above the film's critical temperature. For data from both cooldowns, the phase offset $\alpha$ is found to be $<0.01$ radians.

\section{London-Maxwell modeling}
\label{sec:london-maxwell}

Throughout the measurements, the niobium SQUID susceptometer is kept far below its critical temperature, and the current through the field coil kept far below the critical current of the field coil. In this regime, each layer of the multi-layer superconducting structure of the SQUID responds linearly and non-dissipatively to the applied field coil current, as described by the second London equation,
\begin{equation}
    \label{eq:london-3d}
    \mathbf{H}(\mathbf{r})=-\lambda^2\nabla\times\mathbf{J}_s(\mathbf{r}),
\end{equation}
where $\mathbf{H}$ is the magnetic field in the superconductor and $\mathbf{J}_s$ is the supercurrent density. $\mathbf{H}$ is a sum of the applied magnetic field and the magnetic field due to Meissner screening currents. If $\lambda(T)\gg d$, where $d$ is the film thickness, one can re-write Eq.~\ref{eq:london-3d} in a two-dimensional form in terms of the effective magnetic screening length $\Lambda(T)=\lambda^2(T)/d$ and the thickness-integrated sheet supercurrent density $\mathbf{K}_s=d\mathbf{J}_s$:
\begin{equation}
    \label{eq:london-2d}
    \mathbf{H}(\mathbf{r})=-\Lambda\nabla\times\mathbf{K}_s(\mathbf{r}),
\end{equation}
where we assume that the layer lies parallel to the $x-y$ plane so that $\mathbf{K}_s$ has only $x$- and $y$-components, and only the $z$-component of $\mathbf{H}$ is relevant. Because the sheet supercurrent density is divergenceless, $\nabla\cdot\mathbf{K}_s=0$, we can define a scalar ``stream function'' $g(x, y)$ satisfying
\begin{equation}
    \label{eq:stream}
    \mathbf{K}_s=-\hat{\mathbf{z}}\times\nabla g=\nabla\times(g\hat{\mathbf{z}}).
\end{equation}
This allows us to represent Eq.~\ref{eq:london-2d} as a Poisson equation for $g$:
\begin{equation}
    \label{eq:london-poisson}
    \mathbf{H}(x, y)=\Lambda\nabla^2g(x, y)\hat{\mathbf{z}}.
\end{equation}
We use the open-source Python package \texttt{SuperScreen}~\cite{Bishop-Van_Horn2022-sy} (version 0.8.1~\footnote{\href{https://doi.org/10.5281/zenodo.7796376
}{https://doi.org/10.5281/zenodo.7796376}, \href{https://pypi.org/project/superscreen/0.8.1/}{https://pypi.org/project/superscreen/0.8.1/}}) to solve Eq.~\ref{eq:london-poisson} using the method introduced in Ref.~\cite{Brandt2005-wj}. The method can be applied to 2D films of any shape, including films with holes, and for any value of the effective screening length, $0\leq\Lambda<\infty$.

To calculate the vector potential $\mathbf{A}(\mathbf{r})$ applied to the film by the SQUID susceptometer [Figure~\ref{fig:squid}({\bf d})], we define a model containing the three niobium wiring layers of the SQUID susceptometer [Figure~\ref{fig:squid}({\bf a}, {\bf b})]. We then simulate the response of the entire structure to a current $I_\FC$ flowing in the field coil. The response of all three layers is found self-consistently by iteratively updating the magnetic field applied to each layer based on the Biot-Savart field from the supercurrent flowing in the other layers~\cite{Bishop-Van_Horn2022-sy,Kirtley2016-gt}. Once a self-consistent solution has been found, we can evaluate the magnetic vector potential anywhere in space (in the Lorenz gauge):
\begin{equation}
    \label{eq:A-lorenz}
    \mathbf{A}(\mathbf{r})=\sum_{\text{layers }\ell}\frac{\mu_0}{4\pi}\int_{\ell}\frac{\mathbf{K}_{s,\ell}(\mathbf{r}')}{|\mathbf{r}-\mathbf{r}'|}\mathrm{d}x'\mathrm{d}y',
\end{equation}
where $\mathbf{K}_{s,\ell}$ is the sheet supercurrent density in wiring layer $\ell$, $\mathbf{r}'=(x', y', z_\ell)$ is the position inside the superconductor in layer $\ell$, and $z_\ell$ is the vertical position of the layer. The results of this calculation are shown in Figure~\ref{fig:squid}({\bf d}). The same multi-layer model of the SQUID [Figure~\ref{fig:squid}({\bf a}, {\bf b})] is used to evaluate the flux through the pickup loop $\Phi_\PL$ due to the sheet current density $\mathbf{K}$ flowing in the sample, which is calculated using the TDGL model described in Appendix~\ref{sec:tdgl}. The three niobium wiring layers of the SQUID have $\lambda(T_\mathrm{SQUID})\sim d$, where $T_\mathrm{SQUID}\approx 5\,\mathrm{K}$ is the temperature of the SQUID during the measurement, so they don't satisfy the condition $\lambda(T) \gg d$. Despite this limitation, this 2D London-Maxwell approach has proven effective in modeling the magnetic response of our SQUID susceptometers~\cite{Kirtley2016-gt, Kirtley2016-zz, Bishop-Van_Horn2022-sy}.

\section{Time-dependent Ginzburg-Landau modeling}
\label{sec:tdgl}

To model the nucleation and dynamics of vortices in the niobium film, we use a generalized time-dependent Ginzburg-Landau approach introduced in Refs.~\cite{Kramer1978-kb, Watts-Tobin1981-mn}. The theory is an extension of the time-dependent Ginzburg-Landau theory first developed by Schmid~\cite{Schmid1966-bh} and Gor'kov~\cite{Gorkov1996-do}. The generalized version includes the effect of inelastic electron-phonon scattering, the strength of which is characterized by a parameter $\gamma=2\tau_E\Delta_0$, where $\tau_E$ is the inelastic scattering time and $\Delta_0$ is the zero-field
superconducting gap. This extension makes the theory applicable to gapless superconductors ($\gamma=0$) or dirty gapped superconductors ($\gamma > 0$) where the inelastic diffusion length is much smaller than the coherence length $\xi$~\cite{Kopnin2001-ip}. $\gamma$ essentially characterizes the viscosity for vortex motion in the superconductor~\cite{Jelic2016-ww}, so one would expect $\gamma$ to be relevant to high-frequency viscous flux flow losses, but not to the low-frequency pinning-related losses observed in this work. For the simulations presented in the main text, we set $\gamma=1$.

The model~\cite{Kramer1978-kb,Watts-Tobin1981-mn} boils down to a set of coupled partial differential equations for the complex order parameter of the condensate and the electric scalar potential. In dimensionless units, the equations read:
\begin{equation}
    \label{eq:tdgl-psi}
    \begin{split}
    &\frac{u}{\sqrt{1+\gamma^2|\psi|^2}}\left(\frac{\partial}{\partial t} + i\mu + \frac{\gamma^2}{2}\frac{\partial |\psi|^2}{\partial t}\right)\psi\\
    &=\left(\epsilon-|\psi|^2\right)\psi + (\nabla-i\vec{A})^2\psi
    \end{split}
\end{equation}
\begin{equation}
    \label{tdgl-poisson}
    \begin{split}
        \nabla^2\mu &= \nabla\cdot\mathrm{Im}[\psi^*(\nabla-i\mathbf{A})\psi]\\
        &=\nabla\cdot\mathbf{J}_s
    \end{split}
\end{equation}
$\psi(\mathbf{r}, t)=\Psi(\mathbf{r}, t)/|\Psi_0|$ is the normalized order parameter, where $\Psi_0$ is the zero-field value of the order parameter. $\mu(\mathbf{r}, t)$ is the electric scalar potential, $\mathbf{A}$ is the magnetic vector potential in the superconductor, and $\mathbf{J}_s$ is the supercurrent density. The real-valued parameter $\epsilon(\mathbf{r})=T_c(\mathbf{r})/T - 1 \in [-1,1]$ adjusts the local critical temperature of the film~\cite{Kwok2016-of,Al_Luhaibi2022-cl,Sadovskyy2015-ha}. For all simulations shown here, we fix $\epsilon(\mathbf{r})=1$. Setting $\epsilon(\mathbf{r}) < 1$ suppresses the critical temperature at position $\mathbf{r}$, and extended regions of $\epsilon(\mathbf{r}) < 0$ can be used to model large-scale metallic pinning sites~\cite{Kwok2016-of}. The constant $u=\pi^4/14\zeta(3)\approx5.79$ is the ratio of relaxation times for the order parameter and the magnetic vector potential in dirty superconductors, where $\zeta(x)$ is the Riemann zeta function~\cite{Schmid1966-bh,Kramer1977-oo,Kopnin2001-ip}. Distance is measured in units of the coherence length $\xi=\xi(T)$. The magnetic vector potential $\mathbf{A}$ is measured in units of $A_0=\xi B_{c2}$, where $B_{c2}=\Phi_0/(2\pi\xi^2)$ is the upper critical field. The sheet supercurrent density $\vec{K}_s=d\vec{J}_s$ and sheet normal current density $\vec{K}_n=d\mathbf{J}_n=d\sigma\vec{\nabla}\mu$ are measured in units of $K_0=4\xi B_{c2}/(\mu_0\Lambda)$, where $\sigma$ is the normal state conductivity of the superconductor and $\Lambda=\lambda^2/d$ is the effective screening length. The electric potential $\mu$ is measured in units of 
$V_0=4\xi^2 B_{c2}/(\mu_0\sigma\lambda^2)$, and time is measured in units of $\tau_0=\mu_0\sigma\lambda^2$.

The characteristic time scale $\tau_0=\mu_0\sigma\lambda^2$ is the relaxation time for the magnetic vector potential in the superconductor (or, equivalently, for the current density). The constant $u\approx5.79$ is the ratio of the relaxation times for the order parameter ($\tau_\Psi$) and the vector potential ($\tau_0$) in dirty superconductors~\cite{Schmid1966-bh,Kramer1977-oo,Kopnin2001-ip}. Both of these relaxation times, $\tau_0$ and $\tau_\Psi=u\tau_0$, are several orders of magnitude shorter than the measurement time scale $2\pi/\omega$, justifying the ``quasistatic approximation'' discussed in Section~\ref{sec:modeling}.

Isolating boundary conditions are enforced on superconductor-vacuum or superconductor-insulator interfaces (such as the film edge and the boundary of the slot in Figure~\ref{fig:squid-slot-top}), in the form of Neumann boundary conditions for $\psi$ and $\mu$:
\begin{subequations}
    \label{eq:bc-vacuum}
    \begin{align}
        \hat{\mathbf{n}}\cdot(\nabla-i\mathbf{A})\psi &= 0\label{eq:bc-vacuum-psi}\\
        \hat{\mathbf{n}}\cdot\nabla\mu &= 0\label{eq:bc-vacuum-mu},
    \end{align}
\end{subequations}
where $\hat{\mathbf{n}}$ is a unit vector normal to the interface.

We solve Eqs.~\ref{eq:tdgl-psi} and \ref{tdgl-poisson} on a triangular mesh in two dimensions using the open-source Python package \texttt{pyTDGL}~\cite{Bishop-Van_Horn2023-wr}. The implementation of \texttt{pyTDGL} is based on Refs.~\cite{Jonsson2022-mb, Jonsson2022-xe, Gropp1996-uw, Du1998-kt}. In the analysis presented above, we neglect screening when solving the TDGL model, meaning that we assume the magnetic vector potential in the film is equal to the applied magnetic vector potential, neglecting the induced vector potential due to currents flowing in the film: $\mathbf{A}=\mathbf{A}_\mathrm{applied} + \mathbf{A}_\mathrm{induced}\approx\mathbf{A}_\mathrm{applied}$. Here, $\mathbf{A}_\mathrm{applied}(\mathbf{r})$ is the vector potential from the SQUID susceptometer field coil [Figure~\ref{fig:squid}({\bf c})]. This approximation is easily justified when the effective screening length $\Lambda(T)=\lambda(T)^2/d$ is large compared to either the size of the film or the size of the magnetic source (in our case, the SQUID field coil)~\cite{Tinkham2004-ln, Lemberger2013-lu}. The latter condition is approximately satisfied for the highest temperature at which the simulation was performed, $T=9.37\,\mathrm{K} \approx 0.998\,\Tc$, but will be a source of error at lower temperatures, where $\lambda(T)$ is shorter. It is in principle possible to include screening using \texttt{pyTDGL}, but the added computational cost makes it impractical for the simulations presented in this work.

\bibliography{references.bib}

\end{document}